\documentclass[preprint,floatfix,aps,prd,showpacs,nofootinbib,amsmath,amssymb,amsfonts,superscriptaddress]{revtex4-1}
\usepackage{bm}
\usepackage{graphicx,color}
\usepackage{physics}
\usepackage{dsfont}
\usepackage[normalem]{ulem}
\usepackage{hyperref,url}
\usepackage{mathrsfs}
\usepackage{calrsfs}
\usepackage{bbold}
\usepackage{mathtools}
\usepackage{comment}

\usepackage{subfig}
\usepackage{varioref}
\usepackage[active]{srcltx}

\expandafter\ifx\csname package@font\endcsname\relax\else
\expandafter\expandafter
\expandafter\usepackage
\expandafter\expandafter
\expandafter{\csname package@font\endcsname}%
\fi
\hypersetup{
	colorlinks=true,       
	linkcolor=blue,          
	citecolor=blue,        
}

\usepackage[section]{placeins}
\usepackage{enumitem}
\usepackage{fancybox}
\labelformat{figure}{Fig.~#1}
\def\be {\begin{equation}}
\def\ee {\end{equation}}
\def\bea {\begin{eqnarray}}
\def\eea {\end{eqnarray}}

\def\O  {\Omega}

\DeclareMathAlphabet{\mathcal}{OMS}{cmsy}{m}{n}

\begin{document}
\title{Real-space quantum-to-classical transition of time dependent background fluctuations}

\author{S. Mahesh Chandran} 
\email{maheshchandran@iitb.ac.in}
\affiliation{Department of Physics, Indian Institute of Technology Bombay, Mumbai 400076, India}
\author{Karthik Rajeev} 
\email{karthik\_rajeev@iitb.ac.in}
\affiliation{Department of Physics, Indian Institute of Technology Bombay, Mumbai 400076, India}
\author{S. Shankaranarayanan}
\email{shanki@iitb.ac.in}
\affiliation{Department of Physics, Indian Institute of Technology Bombay, Mumbai 400076, India}
\begin{abstract}
Understanding the emergence of classical behavior from a quantum theory is vital to establishing the quantum origin for the temperature fluctuations observed in the Cosmic Microwave Background (CMB). We show that a real-space approach can comprehensively address the quantum-to-classical transition problem in the leading order of curvature perturbations. To this end, we test spatial bipartitions of quadratic systems for the interplay between three different signatures of classical behavior: i) decoherence, ii) peaking of the Wigner function about classical trajectories, and iii) relative suppression of non-commutativity in observables. We extract these signatures from the covariance matrix of a multi-mode Gaussian state and address them primarily in terms of entanglement entropy and log-classicality. Through a phase-space stability analysis of spatial sub-regions via their reduced Wigner function, we ascertain that the underlying cause for the dominance of classicality signatures is the occurrence of gapped inverted mode instabilities. While the choice of conjugate variables enhances some of these signatures, decoherence studied via entanglement entropy is the stronger and more reliable condition for classicality to emerge. We demonstrate the absence of decoherence, which preempts a quantum-to-classical transition of scalar fluctuations in an expanding background in $(1+1)$-dimensions using two examples: i) a Tanh-like expansion and ii) a de-Sitter expansion. We provide connection between log classicality and particle number by studying the
evolution of each normal mode at late times. We then extend the analysis to leading order fluctuations in $(3+1)-$dimensions to show that a quantum-to-classical transition occurs in the de-Sitter expansion and discuss the relevance of our analysis in distinguishing cosmological models.
\end{abstract}
\maketitle
\section{Introduction}

The emergence of classical behavior of the Universe from its predominantly quantum mechanical early stage is one of the most intriguing phenomena in cosmology~\cite{1989HalliwellPhys.Rev.D,1989PadmanabhanPhys.Rev.D}.
This fascinating process is believed to be rooted in the dynamics shared by generic quantum systems when they interact with their environments. A crucial effect in this context is the loss of quantum coherence induced by the environment. Quantum coherence is a fundamental property of quantum mechanics that results from the superposition of orthogonal states with regard to a reference basis~\cite{2008VedralNature}. Specifically, it refers to the ability of a quantum system to maintain a well-defined quantum state over time, unaffected by external disturbances or interactions. Quantum coherence is necessary for both entanglement~\cite{2010-Eisert.etal-Rev.Mod.Phys.} and other measures of quantum correlations (such as discord, negativity and circuit complexity). It is also vital for quantum computing because quantum algorithms depend on the ability to manipulate and preserve superposition and entanglement. 

Due to the nature of closed quantum evolution, quantum coherence can never vanish permanently from a closed quantum system. However, realistic physical systems are embedded in an inaccessible or partially accessible environment. A quantum system will typically become entangled with many environmental degrees of freedom when interacting with the environment. This entanglement can in turn non-trivially affect local measurements made in the system. Quantum systems progressively lose coherence to the environment due to interactions with the external environment and can be treated as classical~\cite{1982ZurekPhys.Rev.D,1985Joos.ZehZ.Phys.B}. As a closed system, the origin of the classical world requires explanation.


Returning to the cosmological scenario, the Cosmic Microwave Background (CMB)~\cite{2020Aghanim.othersAstron.Astrophys.,2021Alam.othersPhys.Rev.D} provides essential proof of temperature variations in a relatively homogeneous distribution of matter, radiation, and (potentially) dark energy.
These inhomogeneities can be traced all the way back to the early-Universe, and are understood to be seeded by vacuum quantum fluctuations stretched to cosmological scales during a rapidly expanding \textit{inflationary} phase~\cite{1981Mukhanov.ChibisovSovietJournalofExperimentalandTheoreticalPhysicsLetters,1982Guth.PiPhys.Rev.Lett.,1982HawkingPhysicsLettersB,1982StarobinskyPhysicsLettersB,1983Bardeen.etalPhys.Rev.D}. Interestingly, such inhomogeneities, when treated as classical stochastic fluctuations seeded in the CMB after the end of inflation, provide a compelling explanation for the evolution of large-scale structures in the Universe as observed at late-times~\cite{1991Brandenberger.etalPhysicaScripta,1996Polarski.StarobinskyClassicalandQuantumGravity,1998Kiefer.etalInt.J.Mod.Phys.D,2005Lombardo.NacirPhys.Rev.D,2016Martin.VenninPhys.Rev.D}. Then the questions of how vacuum fluctuations evolved to resemble classical fluctuations, and how non-trivial signatures of such a transition can be observed, become pertinent towards establishing the quantum origin of CMB fluctuations~\cite{1990Grishchuk.SidorovPhys.Rev.D,2006Perez.etalClassicalandQuantumGravity,2011SudarskyInt.J.Mod.Phys.D,2020Ashtekar.etalPhys.Rev.D}. 

While there is no single, unified criterion for the emergence of classicality within a quantum field theoretical framework, it is largely addressed via a collection of phenomenological signatures associated with different facets of classical behavior. For instance, as a result of 
 the mixing of the super-Hubble (system) and sub-Hubble (environment) momentum-modes of fluctuations due to non-linear curvature perturbations, the super-Hubble modes are found to \textit{decohere}. A continuously evolving quantum information toolbox comprising of quantum entanglement~\cite{2012Balasubramanian.etalPhys.Rev.D,2017Kumar.ShankaranarayananPhys.Rev.D,2022Choudhury.etalPhys.Rev.D,2020Brahma.etalPhys.Rev.D,2022Brahma.etalClassicalandQuantumGravity,2022Choudhury.etalFortsch.Phys.,2022Choudhury}, quantum discord~\cite{2016Martin.VenninPhys.Rev.D,2022Martin.etalJCAP,2023Martin.etalEurophysicsLetters}, open-effective field theory (EFT) approaches~\cite{2022Brahma.etalJournalofHighEnergyPhysics,2023Burgess.etalJournalofCosmologyandAstroparticlePhysics,2021Banerjee.etal}, in the momentum space has lately proved decisive in making robust predictions for the (extremely rapid) decoherence rate and the (highly suppressed) quantum corrections to the power spectrum resulting from this. However, these signatures are reportedly too small to be captured by current observations. Furthermore, the absence of decoherence in the leading (linear) order of curvature perturbations due to mode-decoupling, and various pitfalls associated with the emergence of classical behaviour in squeezed quantum states have been critically addressed in recent works~\cite{2021Berjon.etalPhys.Rev.D,2022Hsiang.HuUniverse,2022Agullo.etalJournalofCosmologyandAstroparticlePhysics}. 

A real-space approach towards understanding quantum-classical transition is much less explored in this context, in spite of providing a more intuitive picture of field entanglement~\cite{1986-Bombelli.etal-Phys.Rev.D,1993-Srednicki-Phys.Rev.Lett.,1995Mueller.LoustoPhys.Rev.D,2004-Calabrese.Cardy-JournalofStatisticalMechanicsTheoryandExperiment,2009-Calabrese.Cardy-JournalofPhysicsAMathematicalandTheoretical,2010-Eisert.etal-Rev.Mod.Phys.} and its underlying connection with the thermodynamic properties of the background space-time~\cite{1997-Mukohyama.etal-Phys.Rev.D,1998-Mukohyama.etal-Phys.Rev.D,2011-Solodukhin-LivingRev.Rel.,2020Chandran.ShankaranarayananPhys.Rev.D}. While this may have much to do with real-space field-entanglement being plagued by UV-divergences, recent works have proposed ways in which the sensitivity to UV-cutoff can be mitigated through field-smearing in disjoint spatial regions~\cite{2021Martin.VenninPhys.Rev.D,2021Martin.VenninJournalofCosmologyandAstroparticlePhysics,2023Agullo.etal} or scaling symmetry arguments~\cite{2023Chandran.ShankaranarayananPhys.Rev.D}. However, as we will show in this work, the biggest advantage of the real-space picture is that it captures phenomenological signatures of quantum-classical transition even up to the linear order of curvature perturbations. Therefore, the resulting quantum corrections are expected to be significantly less suppressed than in the momentum-space picture.

To identify quantum-classical transition in the real space, we test spatial bipartitions of leading order fluctuations for three different signatures of classical behavior --- i) loss of quantum coherence, which allows the system to be well described by a classical statistical ensemble, ii) peaking of the phase-space distribution of the quantum state about classical trajectories, and iii) relative suppression of non-commutativity. While these signatures may jointly manifest in the momentum-space picture for (higher-order) fluctuations propagating in a (near) de-Sitter background, they are in general inequivalent for the broader class of quantum systems~\cite{2020Ashtekar.etalPhys.Rev.D}. Therefore, the exact interplay between these concepts in real space will be relevant not only for early-Universe fluctuations but also for any quantum system with entangled spatial degrees of freedom. In turn, its applications potentially extend to laboratory simulators for time-dependent backgrounds~\cite{2022Viermann.othersNature,2022SanchezKuntz.etalPhys.Rev.D,2022TolosaSimeon.etalPhys.Rev.A} as well as table-top experiments being proposed for detecting ``quantumness" of gravity in the coming years~\cite{2017Marletto.VedralPhys.Rev.Lett.,2021Rijavec.etalNewJ.Phys.,2022Bose.etalPhys.Rev.D,2023Hanif.etal}.

The paper is organized as follows: In Section \ref{sec:cho}, we develop the tools to extract and measure the aforementioned signatures of classicality in time-dependent quadratic systems, in particular, the CHO system, in detail. Through a phase-space stability analysis of Gaussian states, we identify the presence of gapped inverted modes (in the momentum space) of the entire system as the primary trigger for the quantum-to-classical transition of subsystems (in the real space). In Section \ref{sec:1dcosmology}, we demonstrate the absence of quantum-to-classical transition of scalar fluctuations in an expanding background in $(1+1)-$dimensions using two examples --- i) a Tanh-like expansion and ii) a de-Sitter expansion. Section \ref{sec:3dcosmology} extends the analysis to $(3+1)-$dimensions to show that the quantum-to-classical transition occurs in the de-Sitter expansion but not in the Tanh-expansion. In Section \ref{sec:conc}, we discuss the physical interpretation of our results and future directions. Throughout this work, we use metric signature $(+,-,-,-)$ and set $\hbar=c=1$ unless otherwise specified.

\section{Quantum-to-classical transition in Time-dependent oscillators}\label{sec:cho}
In this section, we analyse the signatures of quantum-classical transition in the phase-space representation of quantum states. We begin our analysis with the coupled harmonic oscillator (CHO) system, which serves as a fundamental building block for the lattice-regularized approach to field theory that will be extensively studied in the later sections. The Hamiltonian for such a system is characterized by a frequency $\omega(t)$ and a coupling parameter $\chi(t)$, both of which are arbitrary (smooth, bounded) functions of time: 
\begin{equation}\label{cho1}
\mathscr{H}(t) =\frac{p_1^2}{2 }+\frac{p_2^2}{2 }+\frac{1}{2} \omega^2(t)\left(x_1^2+x_2^2\right)+\frac{1}{2}\chi^2(t)\left(x_1-x_2\right)^2
\end{equation}
Under the transformations $x_{\pm}=(x_1\pm x_2)/\sqrt{2}$, the above Hamiltonian reduces to:
\begin{equation}
	\label{eq:CHO-Hamil02}
	\mathscr{H}(t) =\frac{p_+^2}{2}+\frac{p_-^2}{2}+\frac{1}{2}\omega_+^2(t)x_+^2+\frac{1}{2}\omega_-^2(t)x_-^2,
\end{equation}
where the time-dependent normal modes are:
\begin{equation}
	\omega_-(t)=\sqrt{\omega^2(t)+2\chi^2(t)}; \qquad  \omega_+(t)=\omega(t).
\end{equation}
We consider the form-invariant Gaussian state (GS), which takes the form~\cite{2008LoheJournalofPhysicsAMathematicalandTheoretical}:
\begin{equation}\label{GS}
\Psi_{\rm GS}(x_+,x_-,t)=\prod_{j=\{+,-\}}\left(\frac{\omega_j(t_0)}{\pi b_j^2(t)}\right)^{1/4}\exp{-\left(\frac{\omega_j(t_0)}{b_j^2(t)}-i\frac{\dot{b}_j(t)}{b_j(t)}\right)\frac{x_j^2}{2}-\frac{i}{2}\omega_j(t_0)\tau_j(t)},
\end{equation}
where $\tau_j =\int b_j^{-2}(t) dt$. The scaling parameters $b_j$ are solutions of the non-linear Ermakov-Pinney equation \cite{Pinney_1950,1967LewisPhys.Rev.Lett.,1968LewisJournalofMathematicalPhysics,2008LoheJournalofPhysicsAMathematicalandTheoretical} :
\begin{equation}\label{ermakov}
	\ddot{b}_j(t)+\omega_j^2(t)b_j(t)=\frac{\omega_j^2(t_0)}{b_j^3(t)}
\end{equation}
The scaling parameters $b_j(t)$ drive the evolution of the Gaussian state as well as its deviation from the initial vacuum state defined at $t=t_0$. While the system evolves to an excited state in the corresponding instantaneous eigenbasis at later time-slices~\cite{2008Mahajan.PadmanabhanGeneralRelativityandGravitation,2018Rajeev.etalGeneralRelativityandGravitation}, its state remains pure $[\Tr\rho^2=1]$ over the course of the evolution. The dynamics of the constituent subsystems ($x_1, x_2$), on the other hand, may exhibit interesting properties by virtue of the entanglement between them. Notably, one subsystem may act as an external environment to the other, causing the latter to ``decohere", or lose some of its quantum features. To illustrate this in the case of CHO, we describe one constituent oscillator (say, $x_2$) with the help of its reduced density matrix (RDM), obtained by tracing out the other oscillator (viz., $x_1$) from the full density matrix of the CHO:

\begin{align}\label{rdm}
	\rho_2(x_2,x_2')&=\int dx_1 \Psi_{\rm GS}^*(x_1,x_2') \Psi_{\rm GS}(x_1,x_2)\nonumber\\
	&=\left(\frac{K_+K_-}{2\pi\Re(A)}\right)^{1/2}\exp{-\frac{\Gamma_1}{2}\left(x_2^2+x_2'^2\right)+\Gamma_2 x_2x_2'+i\frac{\Gamma_3}{2}\left(x_2^2-x_2'^2\right)},
\end{align}
where 
\begin{align}\label{eq:rdmredef}
	\Gamma_1&=2A_R-\left(\frac{B_R^2-B_I^2}{A_R}\right)\quad;\quad
	\Gamma_2=\frac{\abs{B}^2}{A_R}\quad;\quad
	\Gamma_3=2A_I-\frac{2B_RB_I}{A_R}\nonumber\\
 	A&=\frac{1}{4}\left[(K_++K_-)-i(L_++L_-)\right]=A_R+iA_I\nonumber\\
	B&=\frac{1}{4}\left[-(K_+-K_-)+i(L_+-L_-)\right]=B_R+iB_I\nonumber\\
        K_\pm&=\frac{\omega_\pm(t_0)}{b_\pm^2(t)}\quad;\quad
        L_\pm=\frac{\dot{b}_\pm(t)}{b_\pm(t)}
\end{align}
To identify possible signatures of a quantum-classical transition, it is useful to shift to a phase-space representation of the above reduced density matrix.


\subsection{Classicality criteria from phase-space representation}
A phase-space picture is possible within the framework of quantum mechanics with the help of Wigner-Weyl transform~\cite{1987Simon.etalPhys.Rev.A,1987Simon.etalPhysicsLettersA,2008CaseAmericanJournalofPhysics}, which maps operators to phase-space functions:
\begin{equation}
    \mathscr{W}[\hat{O}]\to O(x,p).
\end{equation}
The Wigner-Weyl transform of the density matrix $\rho(x,x')$, also known as the Wigner function, therefore provides a phase-space distribution pertaining to a quantum state:
\begin{equation}\label{eq:WignerGaussian}
    W(x_c,p)=\mathscr{W}[\hat{\rho}]=\frac{1}{2\pi}\int_{-\infty}^{\infty}dx_{\scriptscriptstyle \Delta} \rho\left(x_c-\frac{x_{\scriptscriptstyle \Delta}}{2},x_c+\frac{x_{\scriptscriptstyle \Delta}}{2}\right)e^{-ipx_{_{\scriptscriptstyle \Delta}} },
\end{equation}
where
\begin{equation}
    x_c=\frac{x+x'}{2}\quad;\quad x_{\scriptscriptstyle \Delta}=x-x'.
\end{equation}
For Gaussian states, the Wigner function takes the following form~\cite{1990MorikawaPhys.Rev.D,2008Mahajan.PadmanabhanGeneralRelativityandGravitation}:
\begin{equation}
    W(x,p)=\frac{\alpha}{2\pi\gamma}\exp{-\frac{(p-\beta x)^2}{4\gamma^2}-\alpha^2x^2}.
\end{equation}
In particular, the parameters characterizing the (reduced) Wigner function for the reduced density matrix given in \eqref{rdm}, which shall be our focus in this section, are equated below:
\begin{equation}
    \alpha^2=\Gamma_1-\Gamma_2\quad;\quad \gamma^2=\frac{\Gamma_1+\Gamma_2}{4}\quad;\quad \beta=\Gamma_3,
\end{equation}
where $\Gamma_{1}, \Gamma_{2}$ and $\Gamma_{3}$ are as defined in \eqref{eq:rdmredef}.

The Wigner function is a distribution in the phase-space that exactly captures the probabilistic nature and non-trivial effects (e.g., interference, entanglement) of quantum states in a system, in contrast to well-defined trajectories pertaining to its classical counterpart. The expectation values for observables can be calculated using averages weighted by the Wigner distribution:
\begin{equation}\label{eq:expectation}
    \langle \hat{O} \rangle = \int dx\int dp W(x,p,t)\mathscr{W}[\hat{O}]
\end{equation}
For Gaussian states, the following averages (two-point correlators), computed in the above manner, encode all information about the system:
\begin{equation}\label{eq:averages}
    \langle \{\hat{x},\hat{x}\}\rangle=\frac{1}{\alpha^2}\quad;\quad \langle \{\hat{p},\hat{p}\}\rangle=\frac{\beta^2}{\alpha^2}+4\gamma^2\quad;\quad \langle \{\hat{x},\hat{p}\}\rangle=\frac{\beta}{\alpha^2}
\end{equation}
To better visualize the phase-space features of a Gaussian state, it is convenient to introduce the dimensionless quadratures $P=\frac{p}{\sqrt{2\alpha\gamma}}$ and $X=\sqrt{2\alpha\gamma}x$, in terms of which the Wigner function takes the general form:
\begin{equation} W(X,P)=\frac{\delta_{QD}}{\pi}\exp\left[-\delta_{QD}\left\{\left(P-\frac{1}{\delta_{CC}}X\right)^2+X^2\right\}\right]\quad;\quad 0\leq W\leq \frac{\delta_{QD}}{\pi}
\end{equation}
where $\delta_{QD}$ is referred to as the \textit{degree of quantum decoherence} and 
$\delta_{CC}$ is referred to as the \textit{degree of classical correlations}. The Wigner function is therefore fully characterized by these two dimensionless parameters that capture distinct properties of the quantum state, as outlined below~\cite{1990MorikawaPhys.Rev.D}:
\begin{itemize}
    \item \textbf{Degree of Quantum Decoherence} $\delta_{QD}$ : This measure coincides with the purity of the reduced density matrix $\rho_2$ given in \eqref{rdm}:
    \begin{equation}\label{eq:pur}
        \delta_{QD}\equiv\frac{\alpha}{2\gamma}=\Tr\rho_{2}^2=\sqrt{\frac{4K_+K_-}{(K_++K_-)^2+(L_+-L_-)^2}}
    \end{equation}
Consequently, $\delta_{QD}\in[0,1]$. 
The upper extreme is saturated by the pure states, for which $\delta_{QD}=1$. On the other hand, when the effects of an external environment are significant, the state may undergo decoherence, i.e., the non-diagonal entries drop to zero and the reduced density matrix resembles a classical statistical ensemble. This case corresponds to the limit $\delta_{QD}\to 0$.
    
\item \textbf{Degree of Classical Correlations} $\delta_{CC}$ : This measure is associated with the sharpness of squeezing of the Wigner function:
\begin{eqnarray}\label{eq:dccc}
\delta_{CC}\equiv \abs{\frac{2\alpha\gamma}{\beta}} &=& \frac{\sqrt{K_+K_-\left[(K_++K_-)^2+(L_+-L_-)^2\right]}}{K_+L_-+K_-L_+} \nonumber \\
& = & \frac{2 K_+K_-}{K_+L_-+K_-L_+} \sqrt{\frac{\left[(K_++K_-)^2+(L_+-L_-)^2\right]}{4 K_+K_-}} 
\nonumber \\
&=& \frac{1}{\delta_{QD}} 
\left(\frac{2 K_+K_-}{K_+L_-+K_-L_+}\right)  
\end{eqnarray} 
For the CHO, $\delta_{CC}$ is also directly related to the classicality parameter ($\mathscr{C}$) proposed in \cite{2008Mahajan.PadmanabhanGeneralRelativityandGravitation}. Therein, $\mathscr{C}$ was introduced as a more intuitive measure for quantifying classicality, viz., in terms of the width of the Wigner function around the classical phase-space trajectory. Hence,
    \begin{equation}\label{eq:cpara}
        \mathscr{C}\equiv\frac{\langle xp\rangle_W}{\sqrt{\langle p^2\rangle_W\langle x^2\rangle_W}}=\frac{1}{\sqrt{1+\delta_{CC}^2}}
    \end{equation}
It follows that the classicality parameter $\mathscr{C}\in [0,1]$ (or $\delta_{CC}\in [0,\infty)$). The lower bound corresponds to the ``quantum" limit ($\delta_{CC}\to\infty$) wherein the Wigner function becomes separable in position and momentum. This follows from the uncertainty principle wherein fixing the value of $x$ can amplify the error in $p$ and vice versa, resulting in probability distributions along $x$ and $p$ that are uncorrelated. On the other hand, the upper bound corresponds to the classical limit ($\delta_{CC}\to 0$) wherein the Wigner function is no longer separable in $x$ and $p$, and its peak coincides with well-defined classical phase-space trajectories.
\end{itemize}

As we remarked earlier, there is a convenient geometrical picture that captures the manner in which the above parameters fully characterize a Gaussian state. To visualize this, consider a particular `slice' of the Wigner function that corresponds to an ellipse in the phase space, referred to as a Wigner ellipse, described below in terms of rotated co-ordinates $\tilde{X}$ and $\tilde{P}$:
\begin{align}
    &\frac{\tilde{X}^2}{a^2}+\frac{\tilde{P}^2}{b^2}=\frac{1}{\delta_{QD}}\log{\frac{\delta_{QD}}{\pi W}}\quad;\quad \begin{bmatrix} \tilde{X}\\\tilde{P}\end{bmatrix}=\begin{bmatrix}\cos\theta &\sin\theta\\-\sin\theta&\cos\theta\end{bmatrix} \begin{bmatrix}X\\P\end{bmatrix}\nonumber\\
    &a^2=\frac{1}{b^2}=1+\frac{1}{2\delta_{CC}^2}\left\{1+\sqrt{1+4\delta_{CC}^2}\right\}\nonumber\\
    &\theta=\sin^{-1}\left[\sqrt{\frac{1}{2}\left\{1+\frac{1}{1+4\delta_{CC}^2}\right\}}\right]
\end{align}
where $a$ and $b$ are the lengths of semi-major/minor axes of the rotated ellipse and $\theta$ is the squeezing angle. The squeezing parameter~\cite{1996Polarski.StarobinskyClassicalandQuantumGravity,2023Martin.etalEurophysicsLetters}, which is a popular measure used to characterize squeezed states, can be obtained from the above ellipse as $r=\log\abs{a}$. A useful slice of the Wigner function to look at is at the half of its peak, wherein the corresponding ellipse serves as a 2D-generalization of the FWHM (Full-width at Half-maxima) for Gaussian/normal distributions. The equation for the corresponding Wigner ellipse would then take the form:
 \begin{equation}
    \frac{\tilde{X}^2}{a^2}+\frac{\tilde{P}^2}{b^2}=\frac{\log{2}}{\delta_{QD}}
\end{equation}
When $\delta_{CC}\to\infty$ (or $\mathscr{C}\to0$), we see that the Wigner ellipse reduces to a circle ($a=b=1$, $\theta=\pi/4$) corresponding to a state that is time-independent or at the beginning of its evolution $t=t_0$. This limit also corresponds to zero squeezing ($r\to0$).

The phase-space picture of the quantum state can therefore be outlined as follows: (i) Wigner function for the Gaussian state is fully characterized by dimensionless parameters $\delta_{QD}$ and $\delta_{CC}$ (or $\mathscr{C}$); (ii) State purity $\delta_{QD}$ determines the amplitude features of the Wigner function. For instance, its peak (maxima) and spread (area of the Wigner ellipse at half-maxima) are given by $\delta_{QD}/\pi$ and $\pi\log{2}/\delta_{QD}$ respectively; iii) Classicality parameter $\mathscr{C}$ determines the extent of squeezing ($r=\log{\abs{a}}$) and squeezing angle ($\theta$) of the distribution. From here onwards, we stick to classicality parameter $\mathscr{C}$ as the characteristic measure for squeezing, since it is a fundamental feature of the covariance matrix as we will see in the next subsection, and has a natural extension for large subsystem sizes.

In the phase-space picture, we may now analyze the conditions that must be simultaneously satisfied for classicality to emerge in a Gaussian state~\cite{1990MorikawaPhys.Rev.D,1996Polarski.StarobinskyClassicalandQuantumGravity}:

\begin{itemize}
    \item $\mathscr{C}\to 1$ : In this limit, the Wigner function undergoes a runaway squeezing about the classical phase-space trajectory of the system. 
    \item $\delta_{QD}\to 0$ : In this limit, the subsystem experiences a runaway decoherence due to its interaction with the environment (here, the other oscillator), causing the amplitude of the Wigner function to fall and spread out over the entire phase space. 
\end{itemize}

\begin{figure}
\centering\captionsetup[subfigure]{labelformat=empty}
\subfloat{\label{asd8} \includegraphics[width=0.242\textwidth]{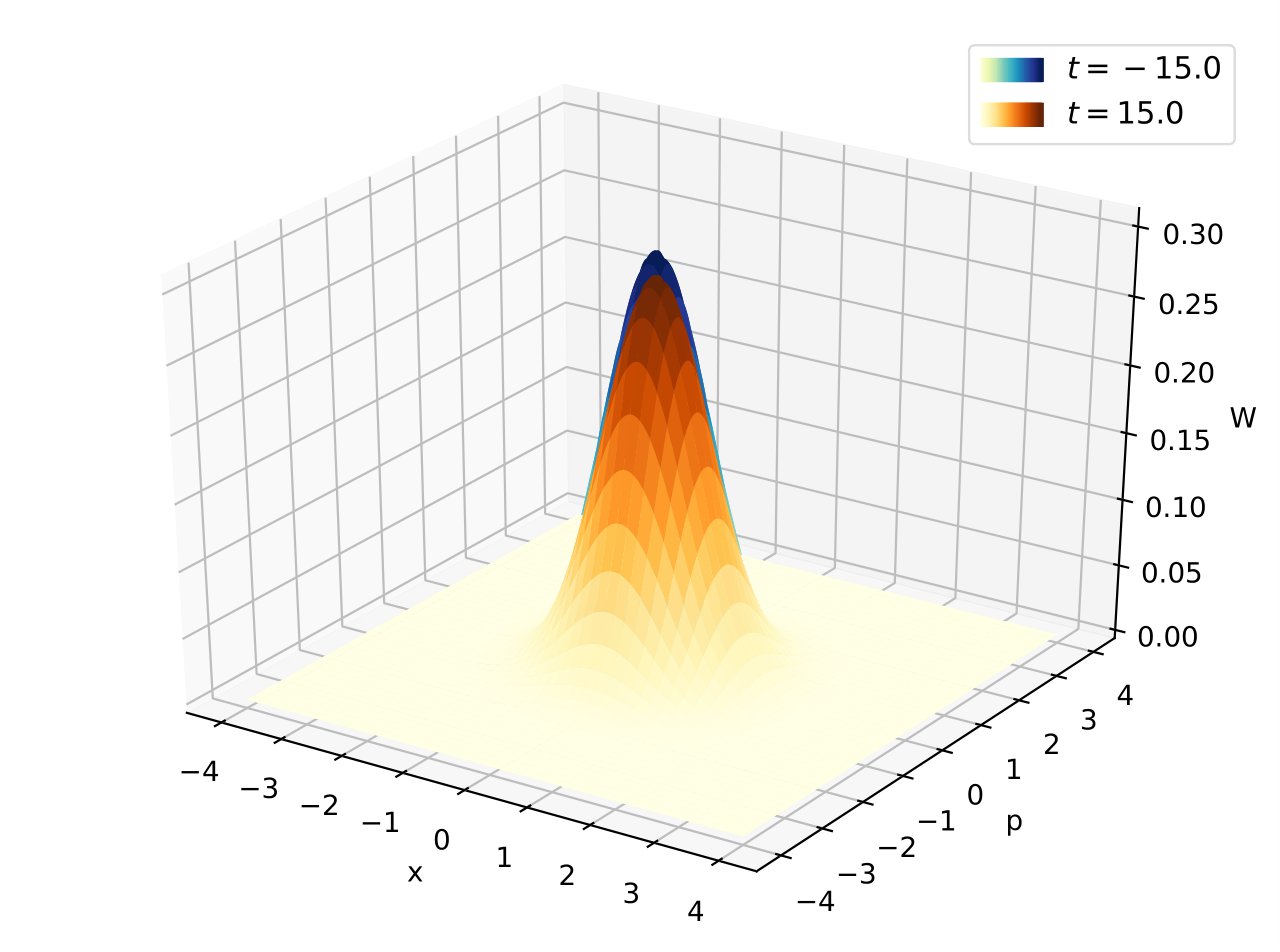}}
\hfill
\subfloat{\label{asd1} \includegraphics[width=0.242\textwidth]{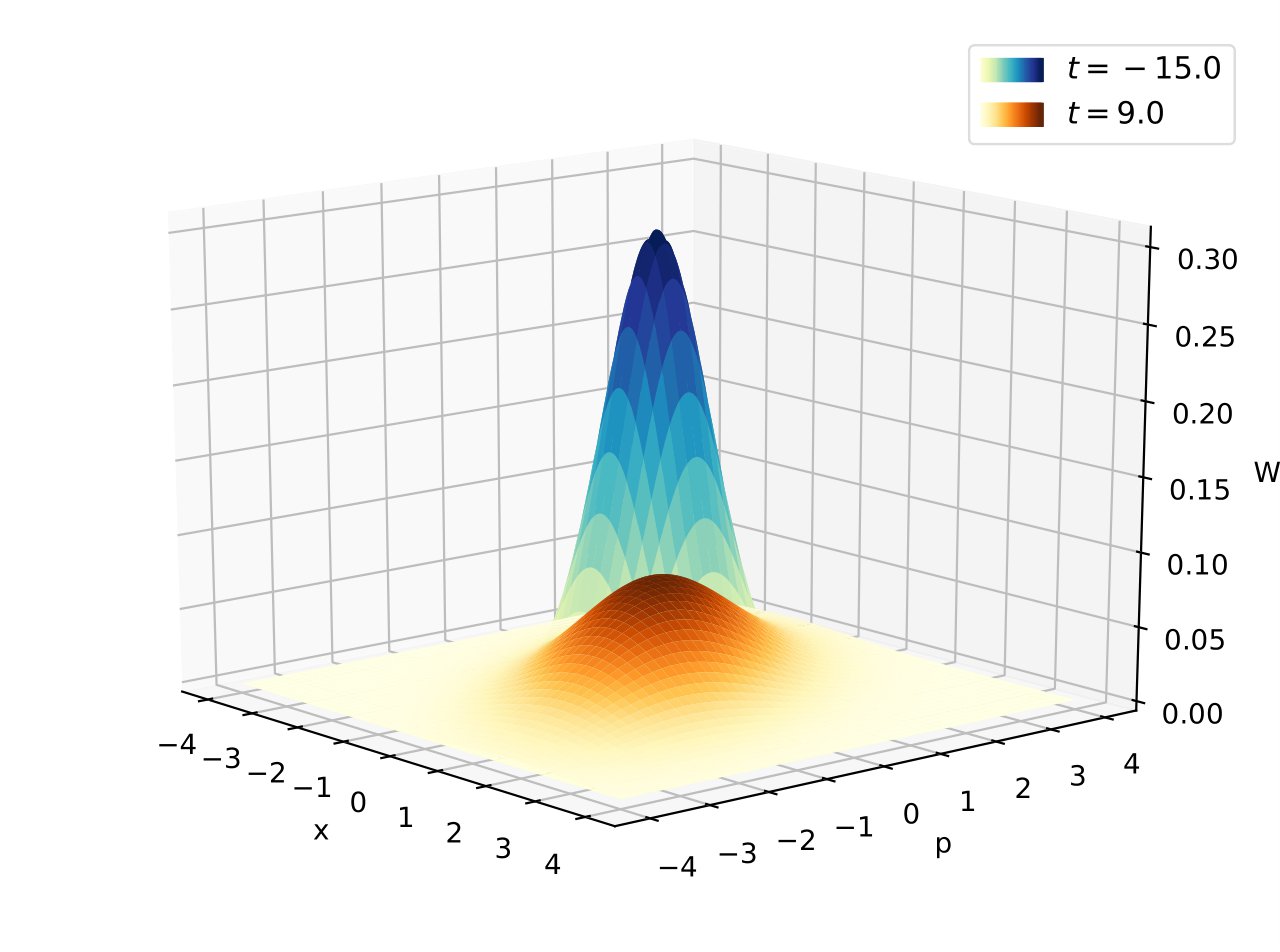}}%
\hfill
\subfloat{\label{asd2} \includegraphics[width=0.242\textwidth]{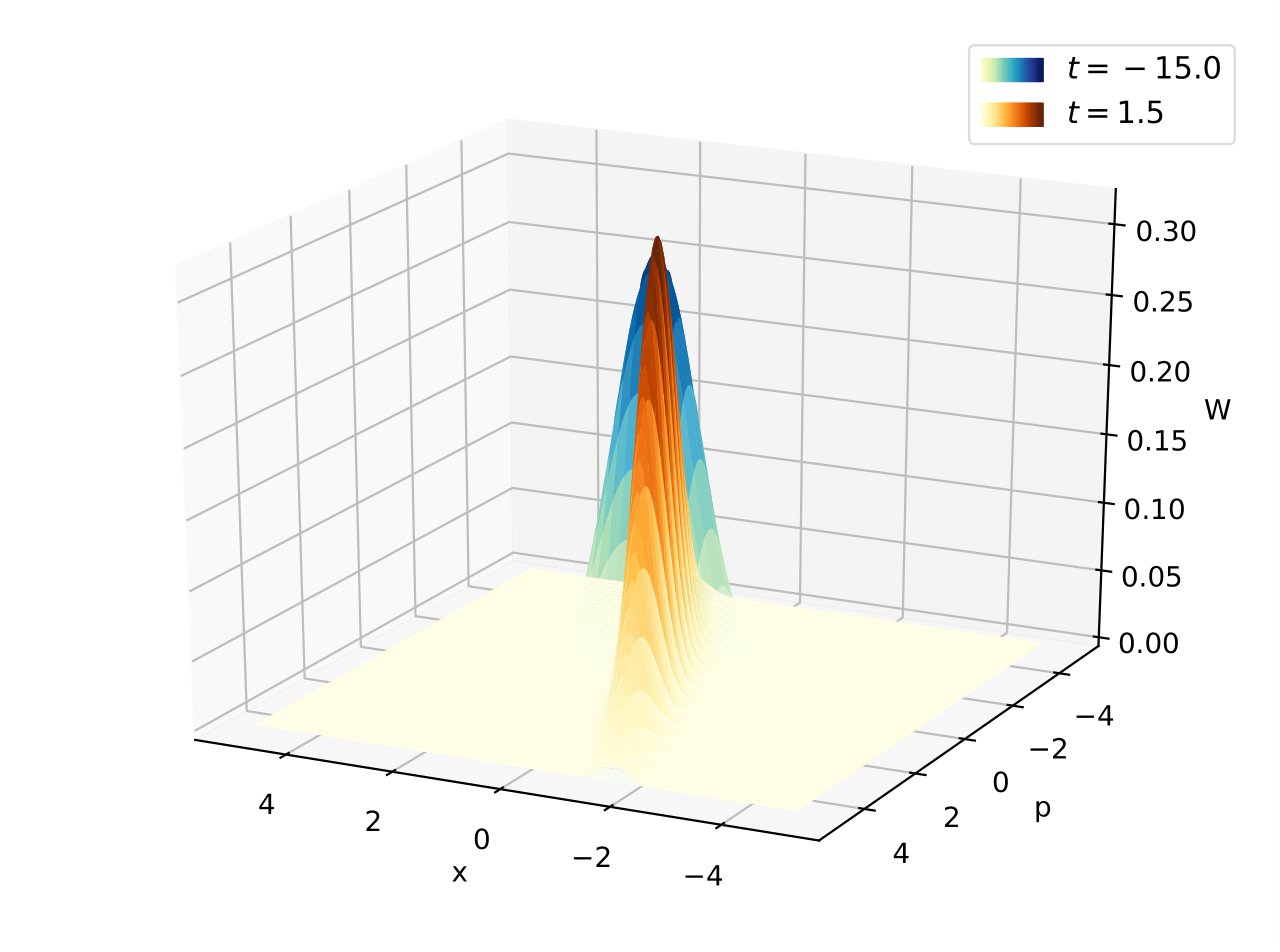}}%
\hfill
\subfloat{\label{asd3} \includegraphics[width=0.242\textwidth]{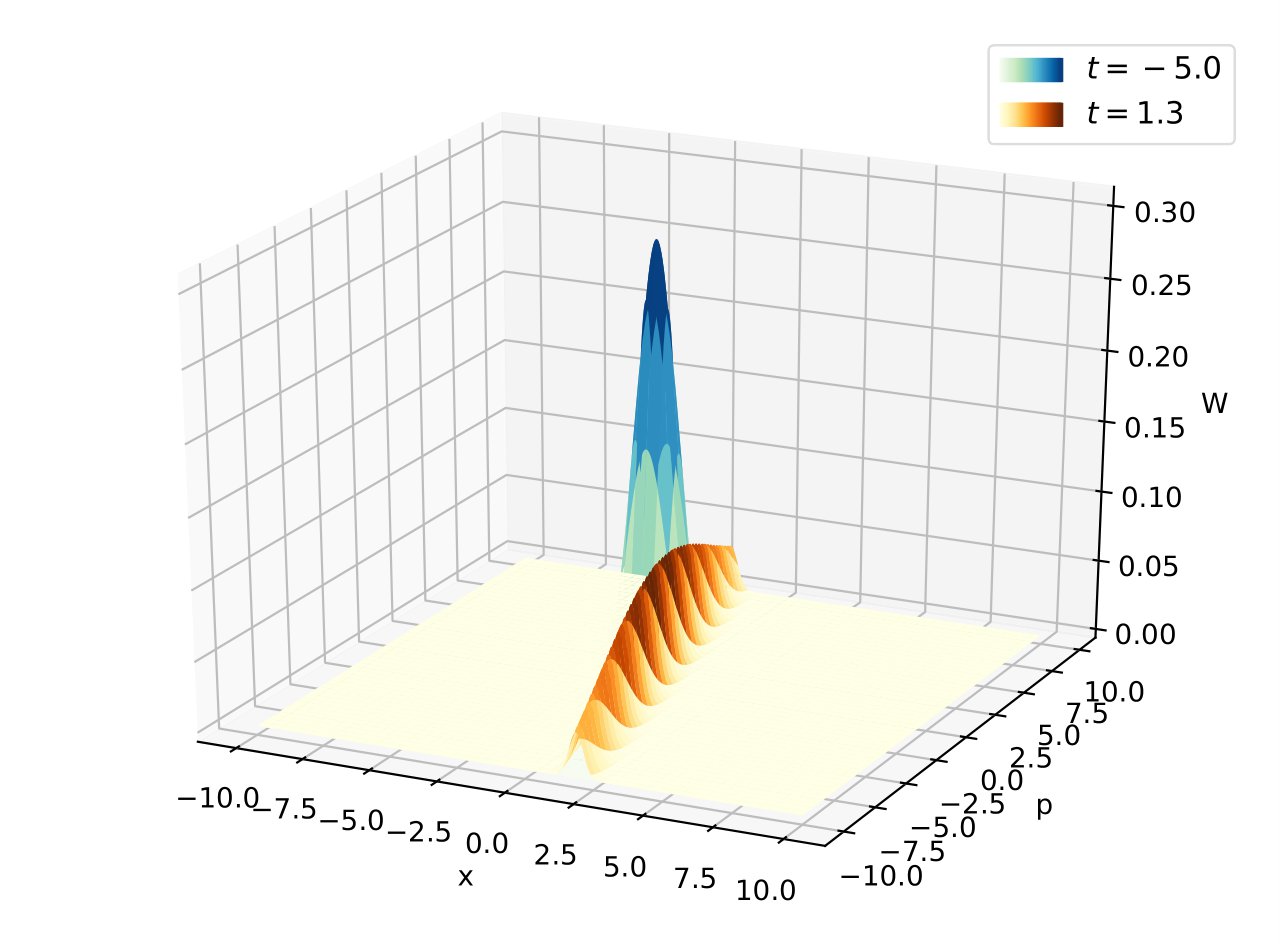}}%
\hfill
\subfloat[(a) $u_\pm^2>0$]{\label{asd4} \includegraphics[width=0.24\textwidth]{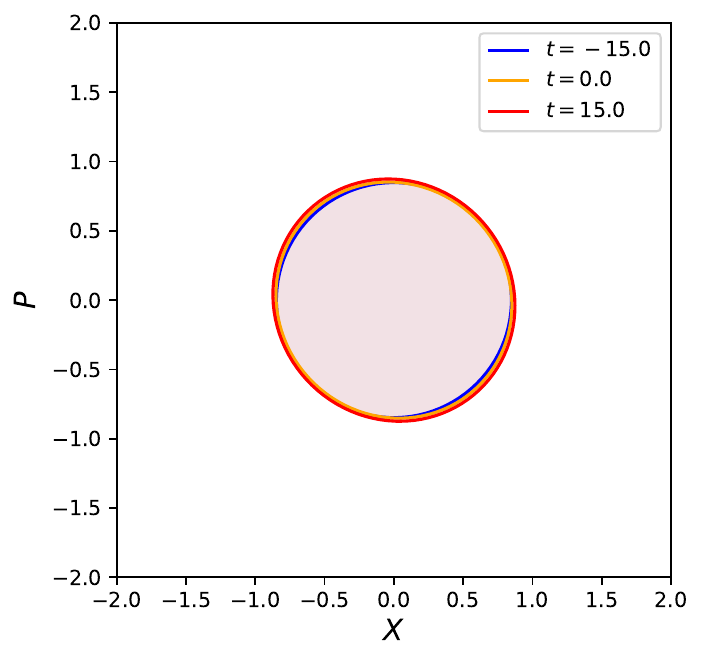}}
\hfill
\subfloat[(b) $u_+^2\to0$]{\label{asd5} \includegraphics[width=0.24\textwidth]{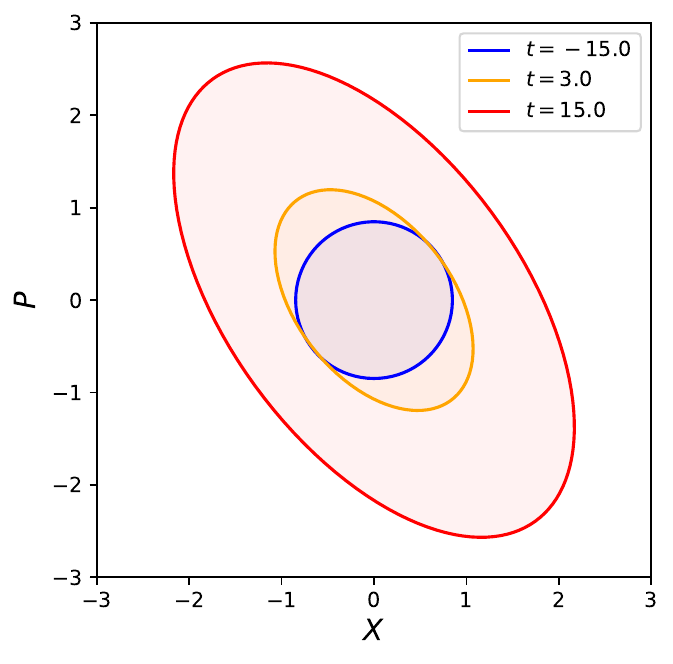}}%
\hfill
\subfloat[(c) $u_\pm^2<0$ (ungapped)]{\label{asd6} \includegraphics[width=0.24\textwidth]{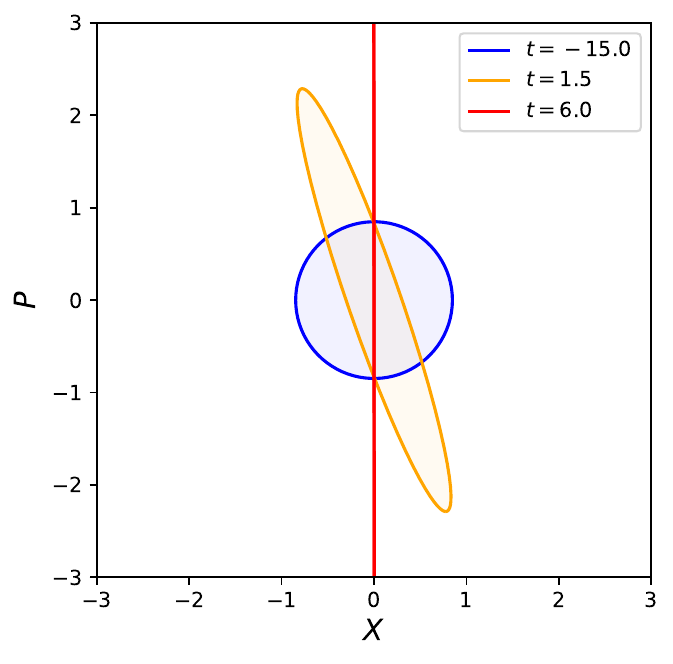}}%
\hfill
\subfloat[(d) $u_\pm^2<0$ (gapped)]{\label{asd7} \includegraphics[width=0.24\textwidth]{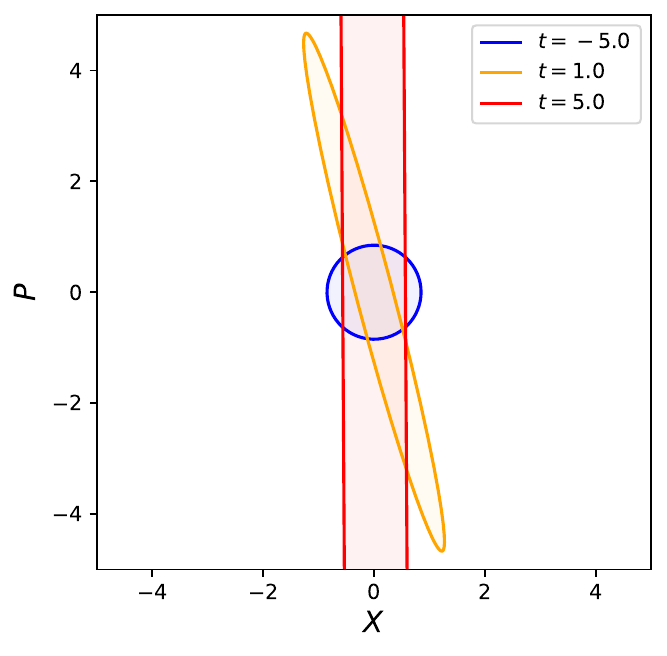}}%
\caption{Evolution of Wigner function (Row 1) and Wigner ellipse at half-maximum (Row 2) for the CHO when $\omega^2(t)$ evolves as \eqref{eq:tanh} with $Q=1$ and $a_0=1$ with constant coupling $\chi(t)=1$ (except for (c)) --- (a) $a_1=0.5$ results in stable modes showing little deviation from the initial ``highly quantum" vacuum state, (b) $a_1=0$ results in a zero mode that decoheres the subsystem but limits the squeezing, (c) $a_1=-0.5$ with $\chi^2(t)=e^{-(t-ti)}$ results in ungapped inverted modes that squeeze the state but limit the decoherence, and (d) $a_1=-2.5$ results in gapped inverted modes that both squeeze and decohere the state, signifying a quantum-classical transition.}
\label{fig:Wigfun}
\end{figure}

To see how the these limits manifest, we perform a phase-space stability analysis in Appendix \ref{sec:ps}, wherein the $k$-mode stability at late-times is ultimately decided by the sign of $u_k^2\equiv\lim_{t\to\infty}\omega_k^2$. The results are summarized in \ref{fig:Wigfun} and Table \ref{tab:CHO}, where we observe that for the CHO, the only case that satisfies the classicality criteria at late-times is when the modes are inverted ($u_{\pm}^2<0$) and gapped ($u_+\neq u_-$). Interestingly, we see that this is also the only regime where entanglement entropy (Appendix \ref{app:cho}) mimics its classical counterpart, the Kolmogorov-Sinai entropy~\cite{2018Hackl.etalPhys.Rev.A,2023Chandran.ShankaranarayananPhys.Rev.D}:
\begin{equation}
    S(t)\equiv-\Tr\rho_{red}(t)\log{\rho_{red}(t)}\sim h_{KS}t\quad;\quad h_{KS}=\sum_i{\lambda_i},
\end{equation}
where growth rate $h_{KS}$ is the sum of all positive Lyapunov exponents. Therefore, we argue that this is indeed the regime where an asymptotic quantum-classical transition occurs in the case of a CHO. While such a criteria has been explored for the CHO to varying extents in previous works~\cite{1996Polarski.StarobinskyClassicalandQuantumGravity,2014Lochan.etalGeneralRelativityandGravitation,2022AndrzejewskiQuantumInformationProcessing}, our approach further reconciles it with the phase-space stability analysis
of the quantum state, in a way that is also scalable to larger subsystem sizes, as we will see in Sec \ref{sec:nho}.

It is to be noted that this notion of ``classicality" fundamentally differs from taking the formal limit $\hbar\to0$~\cite{2022Hsiang.HuUniverse}. To illustrate this, let us briefly put back in the Planck's constant which was set to $\hbar=1$ and consider the Wigner function as well as the  marginal probability distributions along $X$ and $P$ coordinates separately:
\begin{align}
W(X,P)&=\frac{\delta_{QD}}{\pi\hbar}\exp\left[-\frac{\delta_{QD}}{\hbar}\left\{\frac{\tilde{X}^2}{a^2}+\frac{\tilde{P}^2}{b^2}\right\}\right]\quad;\quad \begin{bmatrix} \tilde{X}\\\tilde{P}\end{bmatrix}=\begin{bmatrix}\cos\theta &\sin\theta\\-\sin\theta&\cos\theta\end{bmatrix} \begin{bmatrix}X\\P\end{bmatrix}\nonumber\\
f(X)&=\int dP W(X,P)=\frac{1}{\sqrt{2\pi}\sigma_X}\exp{-\frac{X^2}{2\sigma_X^2}}\quad;\quad\sigma_X=\sqrt{\frac{\hbar}{2\delta_{QD}}}\nonumber\\
g(P)&=\int dX W(X,P)=\frac{1}{\sqrt{2\pi}\sigma_P}\exp{-\frac{P^2}{2\sigma_P^2}}\quad;\quad\sigma_P=\sqrt{\frac{\hbar}{2\delta_{QD}(1-\mathscr{C}^2)}}
\end{align}
It is interesting to note here that while purity $\delta_{QD}$ affects the variance for distributions in both $X$ and $P$, classicality parameter $\mathscr{C}$ only affects the variance in $P$. The measurement error in the $(x,p)$-coordinates is therefore:
\begin{equation}
    \sigma_x\sigma_p=\sigma_X \sigma_P=\frac{\hbar}{2\delta_{QD}\sqrt{1-\mathscr{C}^2}}\geq\frac{\hbar}{2}
\end{equation}
We see that the uncertainty principle is saturated when $\delta_{QD}=1$ and $\mathscr{C}=0$, corresponding to a pure-state at $t=t_0$. The classical limit $\hbar\to 0$, as seen from above, corresponds to the case where the uncertainty (as well as the commutator of conjugate variables) vanishes, and the phase-space distributions are highly localized  ($\delta-$functions)~\cite{1996RipamontiJournalofPhysicsAMathematicalandGeneral,1996Polarski.StarobinskyClassicalandQuantumGravity}:
\begin{equation}
    W\to \delta\left(\frac{\tilde{X}}{a}\right)\delta\left(\frac{\tilde{P}}{b}\right)\quad;\quad f\to\delta(X)\quad;\quad g\to\delta(P)
\end{equation}
Morikawa's classicality criteria on the other hand points to a divergent uncertainty in both $x$ and $p$ measurements, wherein the phase-space distributions are less and less localized (\ref{fig:Wigfun}). Despite this contrast, Morikawa's criteria leads to a notion of ``quasi-classicality"~\cite{1996Polarski.StarobinskyClassicalandQuantumGravity} \textit{within} the framework of quantum mechanics in the following sense --- i) decoherence essentially leads to a (reduced) density matrix that resembles a classical statistical ensemble, and ii) squeezing further aligns the peaks of (reduced) Wigner function along classical phase-space trajectories. The overall implication is that the features that make a state \textit{distinctly} quantum are greatly suppressed. For instance, let us look at the Wigner-Weyl transform of the following observables whose expectation values are to be calculated via \eqref{eq:expectation}:
\begin{align}\label{eq:weyl}
    &\mathscr{W}\left[\hat{x}\hat{p}+\hat{p}\hat{x}\right]= \mathscr{W}\left[2 \mathcal{S}(\hat{x}\hat{p})\right]=2xp\nonumber\\
    &\mathscr{W}\left[\hat{x}^2\hat{p}^2+\hat{p}^2\hat{x}^2\right]= \mathscr{W}\left[2\mathcal{S}(\hat{x}^2\hat{p}^2)+[\hat{x},\hat{p}]^2\right]=2x^2p^2-\hbar^2\nonumber\\
    &\mathscr{W}[f(\hat{x},\hat{p})]=\mathscr{W}\left[\mathcal{S}\left(f(\hat{x},\hat{p})\right)+g\left([\hat{x},\hat{p}]\right)\right]=f(x,p)+\tilde{g}(\hbar),
\end{align}
where $\mathcal{S}$ is a symmetrizer for combinations of $\hat{x}$ and $\hat{p}$ operators, and satisfies $\mathscr{W}\left[\mathcal{S}(\hat{x}^n\hat{p}^m)\right]=x^np^m$~\cite{2005Braunstein.LoockRev.Mod.Phys.,1995Leonhardt.PaulProgressinQuantumElectronics}. The Weyl-transform of a Hermitian, polynomial combination $f(\hat{x},\hat{p})$ of conjugate variables is therefore real-valued phase-space functions that can be split into a ``classical" contribution (from the symmetrizer) and a ``quantum" contribution (from the commutator)~\cite{2008CaseAmericanJournalofPhysics}. Since all higher-order correlators are polynomial functions of two-point correlators for a Gaussian state, it is sufficient to perform a comparison using expectation values of the commutator and the anti-commutator (i.e., symmetrizer at second order):
\begin{equation}\label{eq:commratio}
R_{(x,p)}\equiv\abs{\frac{\langle[\hat{x},\hat{p}]\rangle}{\langle\{\hat{x},\hat{p}\}\rangle}}=\delta_{QD}\delta_{CC}=\frac{\delta_{QD}}{\mathscr{C}}\sqrt{1-\mathscr{C}^2},
\end{equation}
The above ratio compares the strength of quantum and classical contributions over the course of state evolution~\cite{2020Ashtekar.etalPhys.Rev.D}. We see that Morikawa's classicality criteria ($\delta_{QD}\to 0$ and $\mathscr{C}\to1$) leads to an \textit{extremely rapid} suppression of non-commutativity, in favour of the aforementioned notion of quasi-classicality. This further implies that even if we are able to somehow measure observables that directly capture quantum signatures as in \eqref{eq:weyl}, these signatures will be tremendously suppressed by squeezing and/or decoherence, leaving little room to distinguish between a quantum and classical origin for observations. It is also to be noted that the classicality criteria places \textit{stronger} conditions than $R_{(x,p)}\to 0$, requiring simultaneous decoherence ($\delta_{QD}\to0$) \textit{and} squeezing ($\mathscr{C}\to1$). We again set $\hbar=1$ for the rest of the paper, and in the next subsection we will see how the classicality criteria can be extended to a general multi-mode Gaussian state.

\begin{table}[!htb]
	\centering
	\resizebox{0.7\textwidth}{!}{%
		\begin{tabular}{@{}|c|c|c|@{}}
			\toprule
			Asymptotics & $\delta_{QD}\not\to0$ & $\delta_{QD}\to0$ \\  
			\toprule
			$\mathscr{C}\not\to 1$ & Stable modes ($u_\pm^2>0$) & Zero mode ($u_+^2\to0$)  \\
   &$[R_{(x,p)}\not\to0]$ & $[R_{(x,p)}\to0]$\\[6pt] \hline
		$\mathscr{C}\to1$ & Inverted modes ($u_\pm^2<0$) & Inverted modes ($u_\pm^2<0$)\\& Case 1 : $v_+\to v_-$ & Case 2 : $v_+\neq v_-$\\
  &$[R_{(x,p)}\to0]$ & $[R_{(x,p)}\to0]$
   \\[6pt] 
			\toprule
		\end{tabular}
	}
	\caption{Testing classicality criteria for various stability regimes in CHO.}
	\label{tab:CHO}
\end{table}

\subsection{Classicality criteria for $N$ oscillators}\label{sec:nho}

The key to reformulating the classicality criteria for large subsystem sizes lies in the \textit{covariance matrix} of the reduced system. This is because for Gaussian states, all information about correlations are captured in the covariance matrix, which can be effectively used to measure both decoherence as well as squeezing even for large system sizes. In order to see this, let us first write down the Wigner function for a general Gaussian state describing an $m$-oscillator subsystem~\cite{2003Serafini.etalJournalofPhysicsBAtomicMolecularandOpticalPhysics,2012Adesso.etalPhys.Rev.Lett.}:
\begin{align}
    W(\Xi)&=\frac{1}{\pi^m\sqrt{\det\Sigma}}\exp\left[-\Xi^T \Sigma^{-1}\Xi\right]\quad;\quad \Xi\equiv\left\{x_1,..,x_m,p_1,..,p_m\right\},
\end{align}
where the covariance matrix $\Sigma$ is defined as follows~\cite{2010-Eisert.etal-Rev.Mod.Phys.}:
\begin{equation}
    \Sigma=\begin{bmatrix}
        \sigma_{XX}&\sigma_{XP}\\\sigma_{XP}^T&\sigma_{PP}    \end{bmatrix}\,;\, (\sigma_{XX})_{ij}=\langle \{x_i,x_j\}\rangle\,;\,(\sigma_{XP})_{ij}=\langle \{x_i,p_j\}\rangle\,;\,(\sigma_{PP})_{ij}=\langle \{p_i,p_j\}\rangle
\end{equation}
The commutation relations for conjugate variables can be represented as follows:
\begin{equation}
    [\Xi_i,\Xi_j]=i\Omega_{ij}\quad;\quad \Omega=\begin{bmatrix}
        O&\mathcal{I}\\-\mathcal{I}&O\quad
    \end{bmatrix}
\end{equation}
These relations are always preserved via symplectic transformations $M$ that satisfy $M\Omega M^T=\Omega$. The covariance matrix $\Sigma$ can be brought to the Williamson normal form with such a transformation~\cite{1936WilliamsonAmericanJournalofMathematics}:
\begin{equation}
    \tilde{\Sigma}=M\Sigma M^T=\begin{bmatrix}
        diag(\gamma_k)&O\\O&diag(\gamma_k)
    \end{bmatrix}
\end{equation}
The symplectic spectrum $\{\gamma_k\}$ can be obtained from the eigenvalues $\{\pm\gamma_k\}$ of the matrix $i\Omega\Sigma$, and are related to the individual purities as follows~\cite{Ferraro2005}:
\begin{equation}
    \gamma_k=\frac{1}{\delta_{QD}^{(k)}}\quad;\quad \det\tilde{\Sigma}=\prod_{k=1}^m\gamma_k^2 \equiv \frac{1}{\Delta_{QD}^2}
\end{equation}
While the overall purity $\Delta_{QD}=\prod_k \delta_{QD}^{(k)}$ of the reduced state appears to be a natural $m$-oscillator extension of the measure $\delta_{QD}$ in CHO, the \textit{entanglement entropy} of the subsystem is a richer measure of decoherence for larger subsystem sizes~\cite{2003Serafini.etalJournalofPhysicsBAtomicMolecularandOpticalPhysics}. The entanglement entropy for the subsystem from the symplectic eigenvalues as follows:
\begin{equation}\label{eq:EntNHO}
    S=\sum_{k=1}^mS_k\quad;\quad S_k=\left(\frac{\gamma_k+1}{2}\right)\log\left(\frac{\gamma_k+1}{2}\right)-\left(\frac{\gamma_k-1}{2}\right)\log\left(\frac{\gamma_k-1}{2}\right)
\end{equation}
On the other hand, in order to generalize the classicality parameter (measure of classical phase-space correlations) for large subsystem sizes, let us first look at the determinant of the matrix $i\Omega\Sigma$:
\begin{equation}
    \det{i\Omega\Sigma}=(-1)^N\prod{\gamma_k^2}=(-1)^N\det\begin{bmatrix}
        \sigma_{XP}^T&\sigma_{PP}\\-\sigma_{XX}&-\sigma_{XP}    \end{bmatrix}
\end{equation}
Upon resolving the above equation with the help of Schur's complement, we get:
\begin{equation}
Q\equiv\mathcal{I}-\sigma_{XX}^{-1}\sigma_{XP}\sigma_{PP}^{-1}\sigma_{XP}^T\quad;\quad     \det Q = \frac{\det\Sigma}{\det[\sigma_{XX}\sigma_{PP}]}
\end{equation}
We now propose that the classicality parameter for a multi-mode Gaussian state corresponding to $m-$oscillators can be generalized as follows:
\begin{equation}
    \mathscr{C}=\sqrt{1-\det Q}=\sqrt{1- \frac{\det\Sigma}{\det[ \sigma_{XX}\sigma_{PP}]}}
\end{equation}
For the CHO, using \eqref{eq:averages} and \eqref{eq:pur}, the above equation exactly reduces to \eqref{eq:cpara}:
\begin{equation}
    \mathscr{C}=\sqrt{1-\frac{1}{4\langle x^2 \rangle\langle p^2\rangle\delta_{QD}^2}}=\frac{\langle xp\rangle}{\sqrt{\langle x^2\rangle\langle p^2\rangle}}
\end{equation}
The classicality parameter that we have proposed serves as a powerful tool towards quantifying classical correlations in a multi-mode Gaussian state. It effectively captures the relative contribution of the off-diagonal block $\sigma_{XP}$ with respect to the diagonal blocks $\sigma_{XX}$ and $\sigma_{PP}$ in the covariance matrix, i.e., it measures how sharply the multi-variate reduced Wigner function squeezes about classical trajectories. For pure states, it also captures information about particle production due to instabilities (Appendix \ref{app:pp}). For the case of CHO, the above result reduces to \eqref{eq:cpara}. However, for a larger subsystem size, we obtain this measure from the determinant of matrix $Q$. In order to have a better comparison with entanglement entropy of the same subsystem, we further rewrite it in terms of what we refer from here on out as ``log classicality" $LC(t)$:
\begin{equation}
    LC\equiv-\log\sqrt{1-\mathscr{C}^2}=-\frac{1}{2}\log{\left(\det Q\right)},
\end{equation}
The above measure is well-behaved, and is a characteristic feature of a multi-mode covariance matrix. Entanglement entropy and log classicality are therefore insightful single-valued measures that extract the extent of decoherence and squeezing directly from the covariance matrix associated with a given (multi-mode) quantum state. The criteria for asymptotic quantum-classical transition can hence be reformulated for large subsystem sizes as follows:
\begin{equation}\label{eq:criteria}
    \lim_{t\to\infty}S\to\infty \quad;\quad \lim_{t\to\infty} LC\to\infty
\end{equation}
In the above limit, a multi-mode generalization for the ratio defined in \eqref{eq:commratio}
 is also expected to vanish. However, since it is a weaker requirement for classicality than \eqref{eq:criteria}, we do not address such a generalization in this work.
 
 Continuing the phase-space stability analysis for CHO in Appendix \ref{sec:ps}, we see that the inverted modes ($u_{\pm}=iv_\pm$) lead to the following leading order behaviour at late-times, with only the gapped ($v_+>v_-$) case satisfying the classicality criteria:
\begin{equation}
    \lim_{t\to\infty}S\sim \begin{cases}
        (v_++v_-)t &v_+>v_-\\ const. & v_+\to v_-
    \end{cases}\quad;\quad \lim_{t\to\infty}LC\sim \begin{cases}
        (v_+-v_-)t &v_+>v_-\\2v_{\pm}t & v_+\to v_- 
    \end{cases}
\end{equation}
Having successfully generalized the classicality criteria for multi-mode Gaussian states, we may now utilize this to identify quantum-classical transition in physical scenarios modeled by dynamically evolving harmonic lattices. The criteria, however, may have a possible caveat. In general, a scalar field propagating in a background space-time may be quantized in different co-ordinate settings. The respective conjugate variables are related via canonical transformations, and ideally we require a classicality criteria that is independent of the choice of these variables. While a lot of progress has been made in identifying this transition particularly in the two-mode squeezed-state representation in the momentum space~\cite{1996Polarski.StarobinskyClassicalandQuantumGravity,2022Martin.etalJCAP,2023Martin.etalEurophysicsLetters}, the choice of conjugate variables is found to play a crucial role, i.e., a system identified as ``classical" can be made ``quantum" with a simple canonical transformation~\cite{2020Grain.VenninJournalofCosmologyandAstroparticlePhysics}. While We address this in much detail in Appendix \ref{sec:dss}, where we show that entanglement entropy, being a symplectic invariant, is unaffected by canonical transformations, as opposed to log classicality. Therefore, we strengthen the classicality condition in \eqref{eq:criteria} by requiring them to be simultaneously satisfied with respect to two sets of canonical conjugate variables chosen by different lapse functions, failing which an asymptotic quantum-classical transition may be ruled out. Upon improving the classicality criteria this way, we will now proceed to analyze early-Universe fluctuations in the following sections.



		

\section{Early universe fluctuations in $(1+1)-D$}\label{sec:1dcosmology}
In this section, we apply the classicality criteria developed in Section \ref{sec:cho} for fluctuations propagating in an expanding universe in $(1+1)-$dimensions. Although this does not reflect the physical situation that concerns us, the extensive analytic control we have compared to $(3+1)-$dimensions can provide us with valuable insight on how an expanding background affects the ``quantumness" of such fluctuations. The unperturbed FLRW metric in comoving coordinates clocked by cosmic time ($\tilde{t}$) and conformal time ($\tilde{\eta}$) are respectively given below:
\begin{equation}
    ds^2=d\tilde{t}^2-a^2(\tilde{t})d\tilde{x}^2=a^2(\tilde{\eta})\left[d\tilde{\eta}^2-d\tilde{x}^2\right]\quad;\quad d\tilde{t}=a(\tilde{\eta})d\tilde{\eta}
\end{equation}
The action for a massive test scalar field in an arbitrary space-time background is given below:
\begin{equation}
S=\frac{1}{2}\int d\tilde{x}^{\mu} \sqrt{-g}\left[g^{\mu\nu}\partial_{\mu}\tilde{\Phi}\partial_{\nu}\tilde{\Phi} -\tilde{m}_f^2\tilde{\Phi}^2\right]
\end{equation}
In $(1+1)-$dimensions, the above action reduces to~\cite{1992Mukhanov.etalPhys.Rept.}:
\begin{equation}
S= \int d\tilde{t} L~~; \quad  
L= \frac{1}{2} \int d\tilde{x}  \left[\tilde{\Phi}'^2-(\partial_{\tilde{x}}\tilde{\Phi})^2-\tilde{m}_f^2\tilde{\Phi}^2\right]
\end{equation}
Upon defining the canonical momentum as $\tilde{\Pi}=\partial_{\tilde{\Phi}'}L$, and discretizing the system as $\tilde{x}=j\tilde{d}$, we get:
\begin{equation}
    \mathscr{H}[\tilde{\eta}]=\frac{1}{2\tilde{d}}\sum_j\left[\tilde{\Pi}_j^2+\left\{\tilde{\Phi}_j-\tilde{\Phi}_{j+1}\right\}^2+\tilde{d}^2\tilde{m}_f^2a^2(\tilde{\eta})\tilde{\Phi}_j^2\right]=\frac{\mathscr{H}^{(I)}}{\tilde{d}}
\end{equation}
We now absorb the UV cutoff $\tilde{d}$ via appropriate canonical transformations~\cite{2023Chandran.ShankaranarayananPhys.Rev.D}:
\begin{equation}
    \mathscr{H}^{(I)}[\eta]=\frac{1}{2}\sum_j\left[\Pi_j^2+\left\{\Phi_j-\Phi_{j+1}\right\}^2+\Lambda a^2(\eta)\Phi_j^2\right]\quad;\quad \eta=\frac{\tilde{\eta}}{\tilde{d}}\quad;\quad \Lambda=\tilde{d}^2\tilde{m}_f^2,
\end{equation}
where we have now shifted to a Hamiltonian that is fully described by dimensionless conformal time $\eta$ and dimensionless field mass $\Lambda$. When we follow a similar procedure to obtain the Hamiltonian in (dimensionless) cosmic time, we obtain:
\begin{equation}\label{eq:cosmicHam1D}
    \mathscr{H}^{(II)}[t]=\frac{1}{2a(t)}\sum_j\left[\Pi_j^2+\left\{\Phi_j-\Phi_{j+1}\right\}^2+\Lambda a^2(t)\Phi_j^2\right]\quad;\quad t=\frac{\tilde{t}}{\tilde{d}}\quad;\quad \Lambda=\tilde{d}^2\tilde{m}_f^2,
\end{equation}
We see that the two Hamiltonians are connected the same way as worked out in \eqref{eq:hamconnection}:
\begin{equation}
    \mathscr{H}^{(II)}[t]=\frac{\mathscr{H}^{(I)}[\eta(t)]}{a(\eta(t))}
\end{equation}
Following the same procedure as in Appendix \ref{sec:dss}, we finally get:
\begin{equation}
    \mathscr{H}^{(II)}[t]=\frac{1}{2}\sum_j\left[\pi_j^2+\frac{\left(\varphi_j-\varphi_{j+1}\right)^2}{a^2(t)}+\Omega^2(t)\varphi_j^2\right]\quad;\quad \Omega^2(t)=\Lambda+\frac{1}{4}\left(\frac{\dot{a}}{a}\right)^2-\frac{\ddot{a}}{2a}
\end{equation}
It should be noted that on going from conformal-time to cosmic-time Hamiltonian, the regularization that places field amplitudes along the comoving lattice $\tilde{x}=j\tilde{d}$ is preserved. Canonical transformations meanwhile act on the regularized field amplitudes, keeping the lattice structure intact. Therefore any bipartition in the real-space also carries over from $\mathscr{H}^{(I)}(\eta)$ to $\mathscr{H}^{(II)}(t)$, and the spatial entanglement can be directly compared for both representations.  

The normal modes spectrum for $\mathscr{H}^{(II)}(t)$ is given below~\cite{2008-Willms-SIAMJournalonMatrixAnalysisandApplications,2021Jain.etalPhys.Rev.D,2023Chandran.ShankaranarayananPhys.Rev.D}:
\begin{equation}
    \omega_j^2(t)=\Omega^2(t)+\frac{4}{a^2}f_j^2\quad;\quad f_j=\begin{cases}
        \sin\left[\frac{j\pi}{2(N+1)}\right] & \text{Dirichlet} \\
        \sin\left[\frac{(j-1)\pi}{2N}\right] & \text{Neumann}
    \end{cases}
\end{equation}
In the massless limit $\Lambda\to0$, and in terms of dimensionless Hubble paramater $H$, the normal modes become:
\begin{equation}
    \omega_j^2(t)=\frac{4}{a^2}f_j^2-\frac{1}{4}\left(H^2+2\dot{H}\right)\quad;\quad H=\frac{\dot{a}(t)}{a(t)}=\tilde{H}\tilde{d}
\end{equation}
In the thermodynamic limit $N\to\infty$, we may further rewrite the normal mode equation in terms of (dimensionless) co-moving momentum $k_j$ as follows:
\begin{equation}
    4a^2\omega_j^2\sim k_j^2-a^2H^2-2a^2\dot{H}\quad;\quad k_j=\frac{2\pi j}{N}=\tilde{k}_j\tilde{d},
\end{equation}
where we see that the normal mode spectrum maps to Fourier modes (for a lattice this spectrum is just the discrete fourier transform~\cite{2023Choudhury.etalSymmetry}). Shifting from co-moving to physical normal modes ($\bar{\omega}_j=a\omega_j$) and physical momenta ($\bar{k}_j=k_j/a$), we get:
\begin{equation}\label{eq:1dfreq}
    4\bar{\omega}_j^2\sim \bar{k}_j^2-H^2-2\dot{H}
\end{equation}
The normal mode $\bar{\omega}_j$ therefore corresponds to a momentum-mode $\bar{k}_j$ that is either sub-Hubble ($k_j>H$) or super-Hubble ($k_j<H$), whereas its \textit{stability} depends further on $\dot{H}$. We see that the inversion/squeezing of super-Hubble modes in general are amplified by an accelerated expansion ($\dot{H}>0$) and suppressed by a decelerated expansion ($\dot{H}<0$). On the other hand, the stability of sub-Hubble modes is enhanced by a
decelerated expansion ($\dot{H}<0$) and worsened by an accelerated expansion ($\dot{H}>0$).

In the case of $(1+1)$-dimensions, we may easily resolve the problem of quantum-classical transition via the connection formulas developed in Appendix \ref{sec:dss}. In the massless limit $\Lambda\to0$, we observe that the Hamiltonian $H^{(I)}$ is time($\eta$)-independent. As a result of this, the conformal-time scaling parameters are trivially fixed:
\begin{equation}
    B_j(\eta)=1\quad;\quad B_j'(\eta)=0
\end{equation}
Using \eqref{eq:bconnection}, we see that:
\begin{equation}\label{eq:cosmicb}
     \frac{\omega(t_0)}{b^2(t)}=\frac{\Omega(\eta_0)}{a(t)}\quad;\quad \frac{\dot{b}(t)}{b(t)}=\frac{\dot{a}(t)}{2a(t)}
\end{equation}
Since the entanglement entropy is a symplectic invariant, cosmic-time Hamiltonian must also result in constant entropies. However, log classicality depends on the choice of conjugate variables, i.e., in this case, the time co-ordinate employed. In any case, however, since entanglement entropy remains a constant throughout, massless fluctuations  \textit{never} undergo a quantum-classical transition in $(1+1)-$dimensions, regardless of squeezing or the choice of conjugate variables. We confirm this through numerical simulations of the cosmic-time Hamiltonian, which is explicitly time-dependent even in the massless case. For this demonstrative exercise, we consider two types of time-dependent background: (i) $a(t)\propto 1+A \tanh(Qt)$, where $A$ and $Q$ are constants. This describes a universe that smoothly expands by a finite factor over its entire evolution from asymptotic past to future, and (ii) $a(t)\propto e^{H t}$, which corresponds to a De-Sitter universe. 
\subsection{Tanh Expansion}
We first consider a simple evolution used for studying particle-production in an expanding background, with an asymptotic past and future where the in- and out- vacua are well-defined~\cite{1982-Birrell.Davies-QuantumFieldsCurved}:
\begin{equation}\label{eq:tanh}
    a(t)=\frac{1}{2}\left[\left\{a_1+a_0\right\}+\left\{a_1-a_0\right\}\tanh{\left(Qt\right)}\right]=\frac{a_0+a_1e^{2Qt}}{1+e^{2Qt}}
\end{equation}
where $a_0$ and $a_1$ are the respective initial and final values of the evolving scale factor and $Q^{-1}$ is the time-scale of quench. Upon evolving from $t_0\to-\infty$, the scaling parameter for all the modes can be obtained from \eqref{eq:bconnection} as follows:
\begin{equation}
    b_j(t)=\sqrt{\frac{1+\frac{a_1}{a_0}e^{2Qt}}{1+e^{2Qt}}}
\end{equation}

\begin{figure*}[!ht]
	\begin{center}
		\subfloat[\label{Tanh1a}][]{%
			\includegraphics[width=0.4\textwidth]{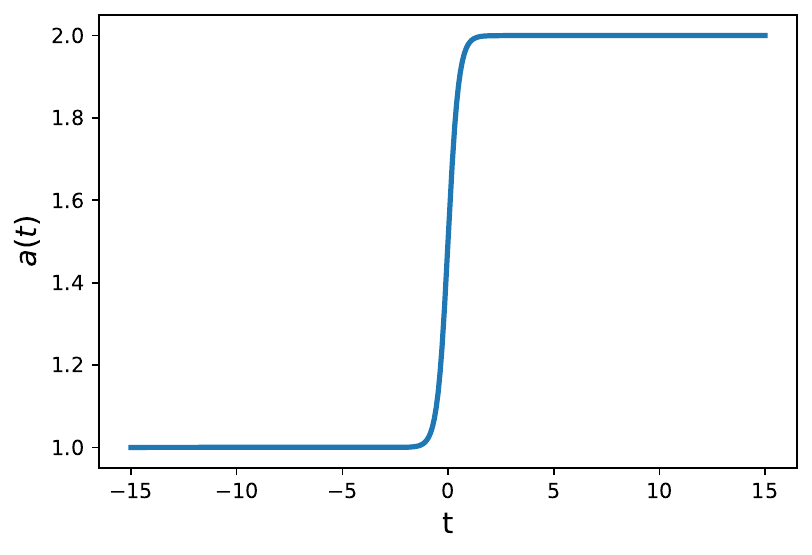}
		}
		\subfloat[\label{Tanh1b}][]{%
			\includegraphics[width=0.4\textwidth]{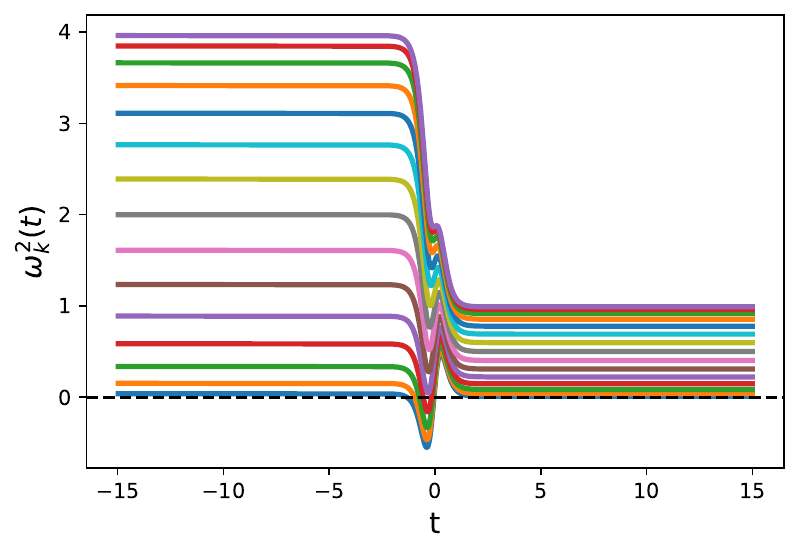}
		}
		
		\caption{Evolution of (a) scale factor $a(t)$ and the corresponding (b) normal mode spectrum in $(1+1)-$dimensions. It can be seen that some normal modes are briefly inverted during the expansion. Here, $N=15$, $a_0=1$, $a_1=2$ and $Q=2$.}
		\label{fig:Tanh1A}
	\end{center}
\end{figure*}
\begin{figure*}[!ht]
	\begin{center}
		\subfloat[\label{Tanh1c}][]{%
			\includegraphics[width=0.4\textwidth]{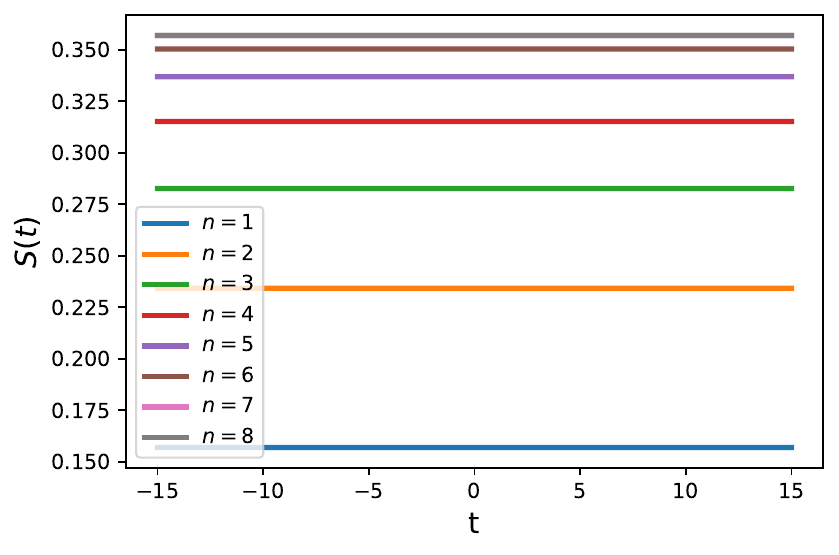}
		}
		\subfloat[\label{Tanh1d}][]{%
			\includegraphics[width=0.4\textwidth]{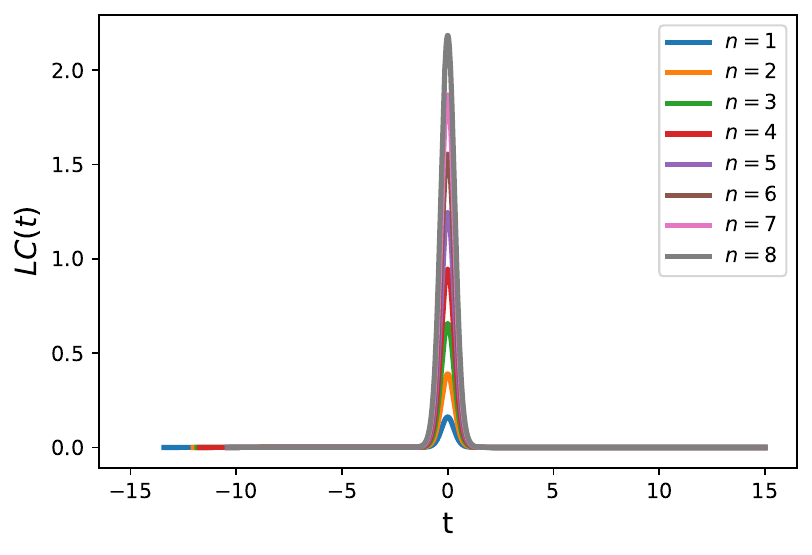}
		}
		
		\caption{Evolution of (a) entanglement entropy $S(t)$ and (b) log classicality $LC(t)$ for a Tanh-quench \eqref{eq:tanh} in $(1+1)-$dimensions. Here, $N=15$ and $H=0.5$.}
		\label{fig:Tanh1B}
	\end{center}
\end{figure*}

From \ref{fig:Tanh1A}, we see that some of the modes undergo a brief inversion during the expansion, signified by the period in which $\omega_{k}^2(t)<0$. While it has no effect on entanglement entropy (symplectic invariance ensures that it stays constant regardless of the choice of conjugate variables), it translates to a brief squeezing of the reduced Wigner function and eventual stabilization, clearly captured by the log classicality plot ($LC$ vs $t$) in \ref{fig:Tanh1B}. This short-lived squeezing is a byproduct of choosing conjugate variables in the cosmic-time Hamiltonian, whereas the same is completely absent ($LC=0$) upon considering conformal-time conjugate variables. Both choices, therefore, fail to satisfy the two-fold classicality criteria.

\subsection{De-Sitter Expansion}
The scale factor during de-Sitter expansion takes the following form:
\begin{equation}\label{eq:dsa}
    a(t)=a_0e^{H(t-t_0)},
\end{equation}
where $a_0$ is the initial value of the scale factor at $t=t_0$, and $H$ is the Hubble constant. Substituting this into \eqref{eq:cosmicHam1D}, we obtain a Hamiltonian that describes a chain of Caldirola-Kanai oscillators~\cite{1941CaldirolaIlNuovoCimento19241942,1948KanaiProgressofTheoreticalPhysics} with nearest-neighbour coupling, and therefore the results we outline here are in turn relevant to understanding dissipative systems. The classical solution for each mode in this case can be obtained by solving:
\begin{equation}
     y_j''(t)+\omega_j^2(t)y_j(t)=0
\end{equation}
where the normal-mode spectrum is given by:
\begin{equation}\label{eq:dsmode}
    \omega_j^2(t)=-\frac{H^2}{4}+\frac{4}{a_0^2}f_j^2e^{-2H(t-t_0)}
\end{equation}
We obtain the independent solutions to be $y_j(t)$ and $y_j^*(t)$, where:
\begin{equation}
    y_j(t)=\exp{\frac{1}{2}Ht+i\frac{2f_j}{a_0H}e^{-H(t-t_0)}}\quad;\quad W[y_j,y_j^*]=4if_j.
\end{equation}
Using \eqref{eq:bsolution}, we obtain the scaling parameters as follows:
\begin{equation}
    b_j^2(t)=e^{H(t-t_0)}\left[1-\frac{a_0H}{4f_j}\sin{\left\{\frac{4f_j(1-e^{-H(t-t_0)})}{a_0H}\right\}}\right]
\end{equation}
It should be noted that the above solution for each $j-$mode is only valid if $a_0H<4f_j$, which along with \eqref{eq:dsmode} tells us that no mode can be inverted at the beginning of the evolution $t=t_0$. In the long-time limit, the scaling parameter takes a similar form as \eqref{b:inverted}:
\begin{equation}
    b_j\sim c_j e^{\frac{H(t-t_0)}{2}}\quad;\quad  c_j=\sqrt{1-\frac{a_0H}{4f_j}\sin{\left(\frac{4f_j}{a_0H}\right)}}
\end{equation}
The key thing to note here is that all the $k-$modes that cross the horizon will have the exact same exponential growth factor ($\sim H/2$) for their respective scaling parameters $b_j(t)$. This eventually results in the saturation of entropy growth (Appendix \ref{app:saturation}), the time-scale ($t_{sat}$) for which is given by the inversion time for the mode with the largest index, i.e., $j=N$. For large $N$, this time-scale can be obtained from \eqref{eq:dsmode}:
\begin{equation}
    t_{sat}\sim t_0+\frac{1}{H}\log{\frac{4}{a_0H}}
\end{equation}

However, if we consider the beginning of the evolution to be at $t_0=-\infty$, the connections and the conditions in \eqref{eq:cosmicb} are satisfied, thereby matching the vacua in both cosmic-time and conformal-time and greatly simplifying the problem. The entanglement entropy therefore saturates instantly ($t_{sat}\to-\infty$) and it remains time-independent throughout the evolution, consistent with the results for the time-independent form conformal-time Hamiltonian. However, the classicality parameter picks up a non-trivial behaviour upon choosing cosmic-time conjugate variables.

\begin{figure*}[!ht]
	\begin{center}
		\subfloat[\label{dS1a}][]{%
			\includegraphics[width=0.4\textwidth]{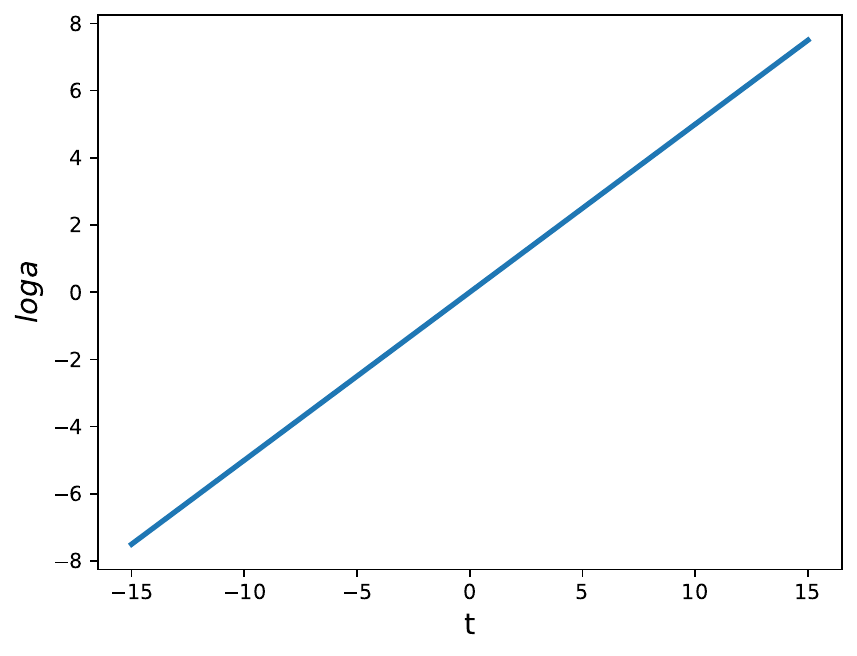}
		}
		\subfloat[\label{dS1b}][]{%
			\includegraphics[width=0.4\textwidth]{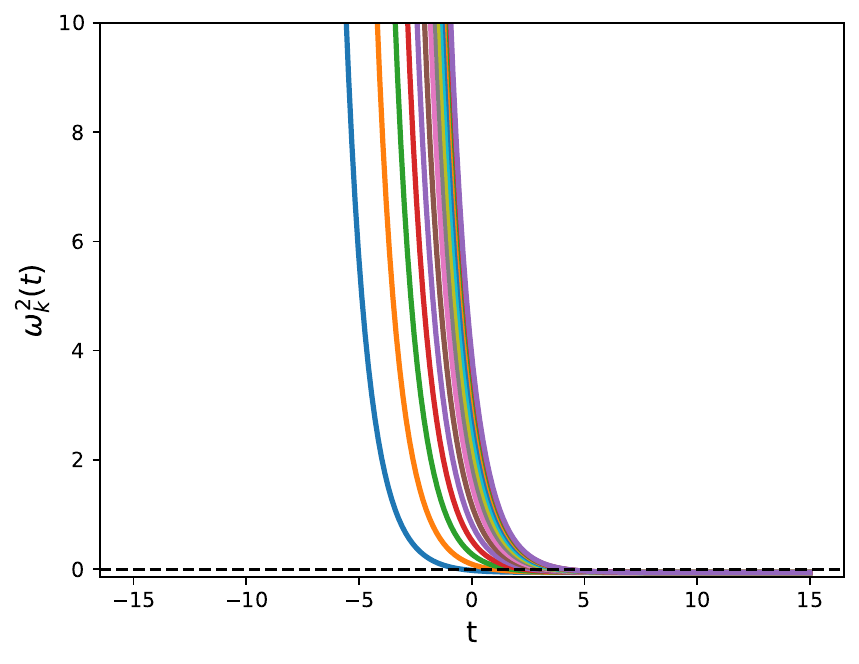}
		}
		
		\caption{Evolution of (a) scale factor $a(t)$ and (b) Normal mode spectrum for de-Sitter expansion \eqref{eq:dsa} in $(1+1)-$dimensions. Here, $N=15$ and $H=0.5$.}
		\label{fig:dS1A}
	\end{center}
\end{figure*}

\begin{figure*}[!ht]
	\begin{center}
		\subfloat[\label{dS1c}][]{%
			\includegraphics[width=0.4\textwidth]{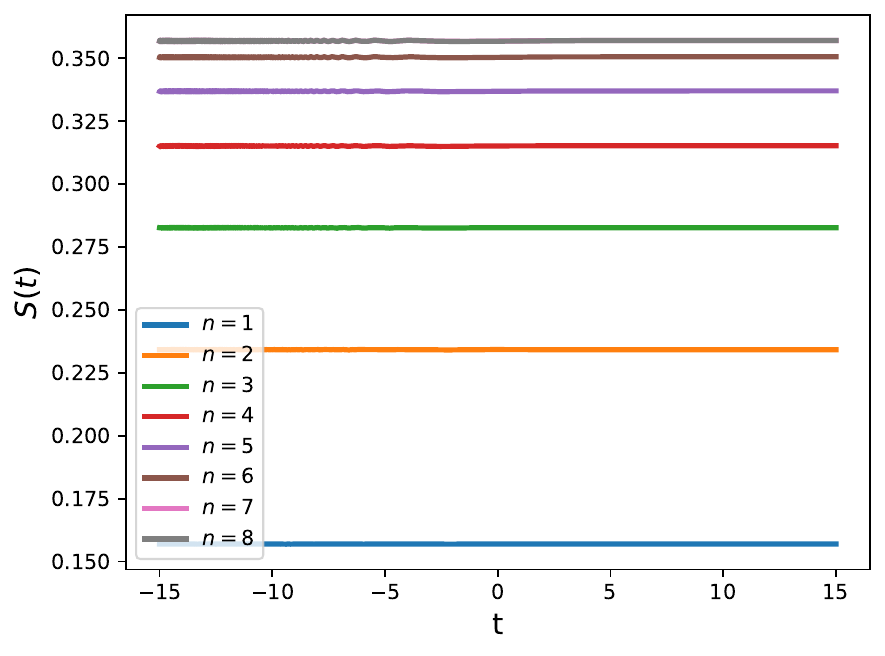}
		}
		\subfloat[\label{dS1d}][]{%
			\includegraphics[width=0.4\textwidth]{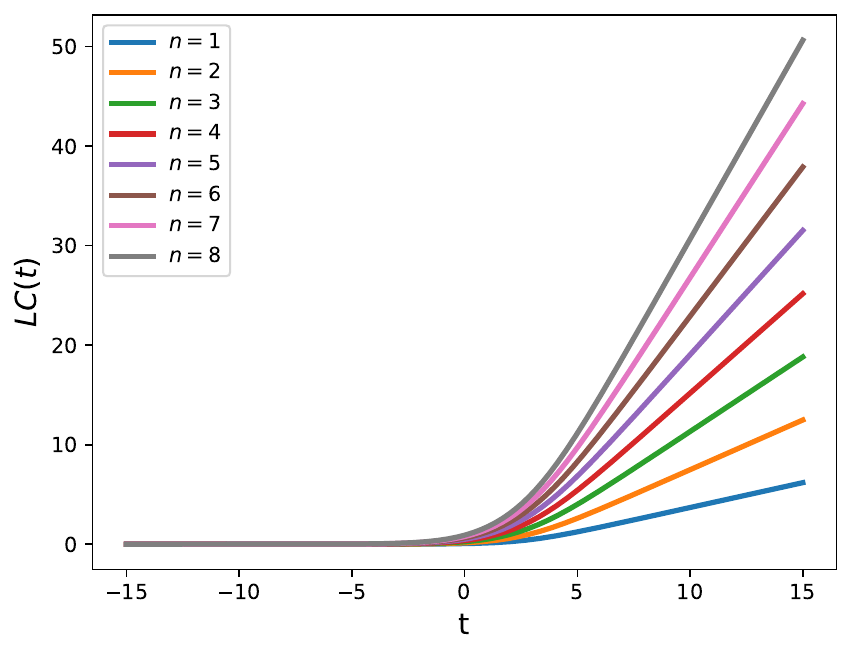}
		}
		
		\caption{Evolution of (a) entanglement entropy $S(t)$ and (b) Log classicality $LC(t)$ for de-Sitter expansion \eqref{eq:dsa} in $(1+1)-$dimensions. Here, $N=15$, $a_0=1$ and $H=0.5$.}
		\label{fig:dS1B}
	\end{center}
\end{figure*}

From \ref{fig:dS1A}, we see that all the normal modes for a de-Sitter expansion in cosmic-time conjugate variables eventually get inverted, and furthermore, they converge asymptotically to the same value, i.e., it exhibits an ungapped inverted mode spectrum as $t\to\infty$. In \ref{fig:dS1B}, the runaway squeezing of the reduced Wigner function translates to a linear growth of log classicality once the first mode becomes inverted (i.e., it has crossed the horizon), and its slope is found to saturate once all the modes have become inverted. The entanglement entropy, despite mode inversion, stays constant. The overall behaviour for any subsystem size in cosmic-time conjugate variables can be summarized below:
\begin{equation}
    S(t)=const\quad;\quad \lim_{t\to\infty} LC(t)\propto Ht
\end{equation}

While the Tanh and de-Sitter models have proved useful in understanding the effects of mode-inversion and squeezing of the Wigner function for large subsystem-sizes, the time-independent behaviour of entanglement entropy effectively rules out any occurrence of quantum-classical transition in $(1+1)-$dimensions. This is also a perfect example of --- i) how squeezing by itself does not imply classicality, echoing the ungapped inverted mode scenario in the CHO (Table \ref{tab:CHO}), and ii) how inverted mode instabilities do not always generate a linear growth in entanglement entropy, in contrast with recent works~\cite{2018Hackl.etalPhys.Rev.A,2023Chandran.ShankaranarayananPhys.Rev.D}. We now turn our attention to $(3+1)-$dimensions in the next section, and apply the classicality criteria for fluctuations propagating in an expanding background.

\section{Early-universe fluctuations in $(3+1)-$D}\label{sec:3dcosmology}
In this Section, we apply the classicality criteria for fluctuations propagating in $(3+1)-$dimensions. The unperturbed expanding background in $(3+1)-$dimensions in co-moving coordinates ($\tilde{r}$,$\theta$,$\phi$) clocked by cosmic time ($\tilde{t}$) or conformal time ($\tilde{\eta}$) is described by:
\begin{equation}
    ds^2=d\tilde{t}^2-a^2(\tilde{t})(d\tilde{r}^2+\tilde{r}^2d\Omega^2)=a^2(\tilde{\eta})\left[d\tilde{\eta}^2-(d\tilde{r}^2+\tilde{r}^2d\Omega^2)\right]\quad;\quad d\tilde{t}=a(\tilde{\eta})d\tilde{\eta},
\end{equation}
where $d\Omega^2=d\theta^2+\sin^2{\theta}d\phi^2$. In terms of the conformal-time, the Lagrangian for a massive test scalar field in an expanding background is given by:
\begin{equation}
    L=\frac{a^2(\tilde{\eta})}{2}\int d\tilde{r}d\theta d\phi \tilde{r}^2\sin{\theta}  \left[\tilde{\Phi}'^2-(\partial_{\tilde{r}}\tilde{\Phi})^2-\frac{1}{\tilde{r}^2}(\partial_{\theta}\tilde{\Phi})^2-\frac{1}{\tilde{r}^2\sin^2{\theta}}(\partial_{\phi}\tilde{\Phi})^2-a^2(\tilde{\eta})\tilde{m}_f^2\tilde{\Phi}^2\right]
\end{equation}
In the massless limit, the above system equivalently describes the leading (linear) order scalar perturbations of the background metric. We may employ spherical decomposition to reduce the system to an effective $(1+1)-$dimensional system~\cite{1993-Srednicki-Phys.Rev.Lett.,2020Chandran.ShankaranarayananPhys.Rev.D}:
\begin{equation}
    \tilde{\Pi}=\frac{1}{\tilde{r}}\sum_{lm}\tilde{\Pi}_{lm}(\tilde{r})Z_{lm}(\theta,\phi)\quad;\quad \tilde{\Phi}=\frac{1}{\tilde{r}}\sum_{lm}\tilde{\Phi}_{lm}(\tilde{r})Z_{lm}(\theta,\phi)
\end{equation}
Upon further obtaining the canonical momentum $\tilde{\Pi}_{lm}=\partial_{\tilde{\Phi}_{lm}'}L$, and discretizing the system as $\tilde{r}=j\tilde{d}$, we get:
\begin{equation}
    \mathscr{H}[\tilde{\eta}]=\frac{1}{2\tilde{d}}\sum_{lmj}\left[\frac{\tilde{\Pi}_{lmj}^2}{a^2(\eta)}+a^2(\eta)\left(j+\frac{1}{2}\right)^2\left\{\frac{\tilde{\Phi}_{lmj}}{j}-\frac{\tilde{\Phi}_{lm,j+1}}{j+1}\right\}^2+\tilde{d}^2\tilde{m}_f^2a^4(\tilde{\eta})\tilde{\Phi}_{lmj}^2\right]=\sum_{lm}\frac{\mathscr{H}_{lm}^{(I)}}{\tilde{d}}
\end{equation}
The UV-cutoff $\tilde{d}$ can be absorbed and the Hamiltonian can be rewritten in terms of dimensionless parameters as follows~\cite{2020Chandran.ShankaranarayananPhys.Rev.D}:
\begin{equation}\label{eq:3dheta}
    \mathscr{H}_{lm}^{(I)}[\eta]=\frac{1}{2}\sum_{lmj}\left[\Pi_{lmj}^2+\left(j+\frac{1}{2}\right)^2\left\{\frac{\Phi_{lmj}}{j}-\frac{\Phi_{lm,j+1}}{j+1}\right\}^2+\left(\Lambda a^2(\tilde{\eta})-\frac{a''(\eta)}{a(\eta)}+\frac{l(l+1)}{j^2}\right)\Phi_{lmj}^2\right]
\end{equation}
where we have defined dimensionless conformal time $\eta=\tilde{d}^{-1}\tilde{\eta}$ and dimensionless field mass $\Lambda=\tilde{d}^2\tilde{m}_f^2$. Unlike the $(1+1)$-D case, we see that the massless conformal-time Hamiltonian has an explicit time($\eta$)-dependence in $(3+1)-D$ case, thereby leading to non-trivial dynamics in entanglement entropy and log classicality. Note that each $l$ mode is independent and hence their contributions to the entanglement entropy can be summed~\cite{1993-Srednicki-Phys.Rev.Lett.,2020Chandran.ShankaranarayananPhys.Rev.D} When we follow a similar procedure to obtain the Hamiltonian in cosmic time, we see that the following relation holds as laid out in Section \ref{sec:dss}:
\begin{equation}
    \mathscr{H}_{lm}^{(II)}[t]=\frac{\mathscr{H}_{lm}^{(I)}[\eta]}{a(\eta(t))}.
\end{equation}
We therefore obtain the following Hamiltonian in terms of (dimensionless) cosmic time:
\begin{equation}
    \mathscr{H}_{lm}^{(II)}[t]=\frac{1}{2}\sum_{j}\left[\Pi_{lmj}^2+\frac{1}{a^2(t)}\left(j+\frac{1}{2}\right)^2\left\{\frac{\Phi_{lmj}}{j}-\frac{\Phi_{lm,j+1}}{j+1}\right\}^2+\Omega_{lmj}^2(t)\Phi_{lmj}^2\right]\quad;\quad t=\frac{\tilde{t}}{\tilde{d}},
\end{equation}
where,
\begin{equation}\label{eq:diag3d}
    \Omega_{lmj}^2(t)=\Lambda+\frac{l(l+1)}{j^2a^2(t)}-\frac{3}{4}\left(\frac{\dot{a}(t)}{a(t)}\right)^2-\frac{3\ddot{a}(t)}{2a(t)}
\end{equation}
Unlike in $(1+1)$-dimensions, the coupling matrix in $(3+1)$-dimensions is not a Toeplitz matrix, as a result of which an exact analytic expression for the normal mode spectrum cannot be obtained~\cite{2020Chandran.ShankaranarayananPhys.Rev.D}. However, we can greatly simplify the problem by splitting the coupling matrix $K$ as follows:
\begin{equation}
    K=\left[\Lambda-\frac{3}{4}\left(\frac{\dot{a}(t)}{a(t)}\right)^2-\frac{3\ddot{a}(t)}{2a(t)}\right]\mathcal{I}+\frac{1}{a^2(t)}\tilde{K},
\end{equation}
where the non-zero elements of $\tilde{K}$ are given below:
\begin{equation}
    \tilde{K}_{jj}=\frac{l(l+1)+\frac{1}{2}}{j^2}+2\quad;\quad \tilde{K}_{j,j+1}=\tilde{K}_{j+1,j}=-\frac{\left(j+\frac{1}{2}\right)^2}{j(j+1)}
\end{equation}
It is easy to see that the matrix that diagonalizes the time-independent, $l$-dependent $\tilde{K}$-matrix also diagonalizes the time-dependent coupling matrix $K$. The normal modes can therefore be written as follows:
\begin{equation}\label{eq:nm3d}
    \omega_{j,l}^2(t)=\Lambda-\frac{3}{4}\left(\frac{\dot{a}(t)}{a(t)}\right)^2-\frac{3\ddot{a}(t)}{2a(t)}+\frac{F_j^2(l)}{a^2(t)},
\end{equation}
where $F_j^2$ are the eigenvalues of $\tilde{K}$, and are therefore also time-independent. While the exact analytical expression for $F_j^2(l)$ cannot be easily calculated, it can be treated as a constant parameter in solving the time-evolution of the scaling parameters $b(t)$ for all modes.

Similar to what was observed in (1+1)-dimensions \eqref{eq:1dfreq}, we can infer that mode inversion is facilitated in cases of accelerated expansion, i.e., $\ddot{a}>0$. The extra input that we get in $(3+1)$ is that the $l$-dependent term $a^{-2}F_j^2$, whose contribution is maximum in the early stages, \textit{counters} mode-inversion. Since $F_j^2$ increases monotonically with $l$, the low-$l$ modes are the first to get inverted, whereas large-$l$ modes follow suit at later times. Since the angular momentum modes are independent, we sum their individual contributions, which are expected to converge as $l\to\infty$ for $(3+1)$-dimensions~\cite{1993-Srednicki-Phys.Rev.Lett.}, as follows:
\begin{equation}
    S(t)=\sum_l (2l+1)S_l(t)\quad;\quad LC(t)=\sum_l (2l+1)LC_l(t)
\end{equation}
For the rest of this section, we rely on numerics to see how various expansion models fare in the classicality test developed in Section \ref{sec:cho}.
\subsection{Tanh Evolution}
\begin{figure*}[!ht]
	\begin{center}
		\subfloat[\label{Tanh2a}][]{%
			\includegraphics[width=0.4\textwidth]{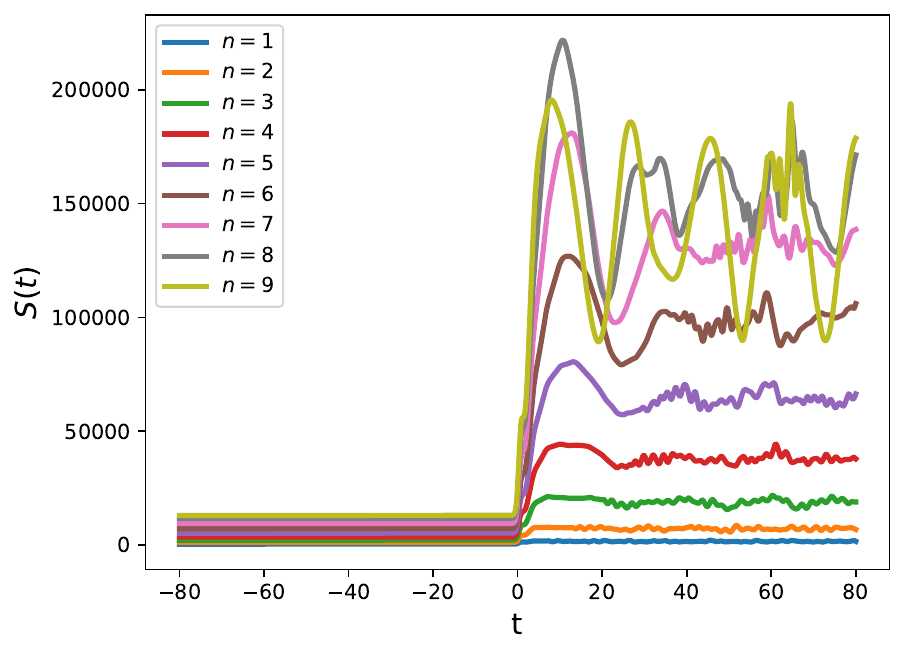}
		}
		\subfloat[\label{Tanh2b}][]{%
			\includegraphics[width=0.4\textwidth]{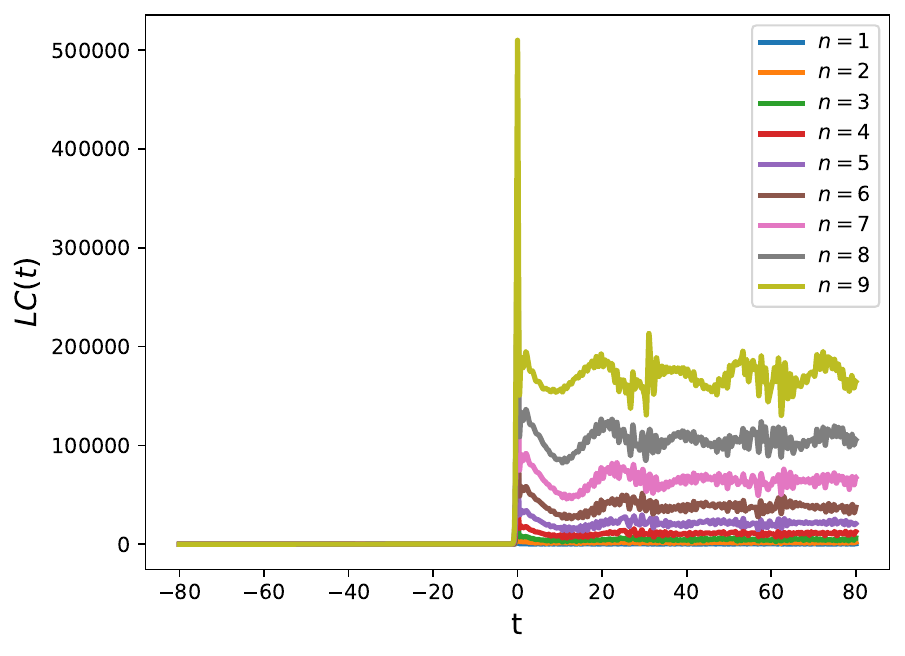}
		}
		
		\caption{Evolution of (a) entanglement entropy $S(t)$ and (b) Log classicality $LC(t)$ for Tanh expansion \eqref{eq:tanh} in $(3+1)-$dimensions. Here, $a_0=1$, $a_1=2$, $Q=2$, $N=10$ and we count up to $l=3000$.}
		\label{fig:Tanh2}
	\end{center}
\end{figure*}

For the same quench function used in \eqref{eq:tanh}, we see from \ref{fig:Tanh2} that both entanglement entropy and log classicality relax to a stable oscillatory behaviour at late-times after an initial surge when the expansion kicks in. Furthermore, unlike the behaviour observed in $(1+1)-$dimensions, the entanglement entropy is no longer time-independent, and log classicality does not revert back to zero at late-times. The latter further indicates particle-production at late-times resulting from the expansion (Appendix \ref{app:pp}), in contrast with the results from $(1+1)-$dimensions.

The above results arise from the fact that during a Tanh expansion in $(3+1)-$dimensions, the momentum modes \eqref{eq:nm3d} of discretized linear fluctuations briefly become inverted when they cross the Hubble radius, and stabilize when they reenter. Such an evolution causes \textit{spatial} bipartitions of fluctuations in the co-moving frame to decohere ($S$ increases) and develop classical correlations (sharp peak in $LC$) during this brief inversion. Upon reentering, although further classicalization is averted, there is an irreversible loss of quantum coherence along with gain in classical correlations resulting from the expansion. This also implies that $R_{(x,p)}$, the relative strength \eqref{eq:commratio} of quantum to classical contributions in observables, is also irreversibly suppressed to some extent, depending on the parameters of the expansion. Therefore, for Tanh expansion, an asymptotic quantum-to-classical transition of fluctuations is avoided in both $(1+1)-$ and $(3+1)-$dimensions.

\subsection{de-Sitter Expansion}
\begin{figure*}[!ht]
	\begin{center}
		\subfloat[\label{dS2aa}][]{%
			\includegraphics[width=0.4\textwidth]{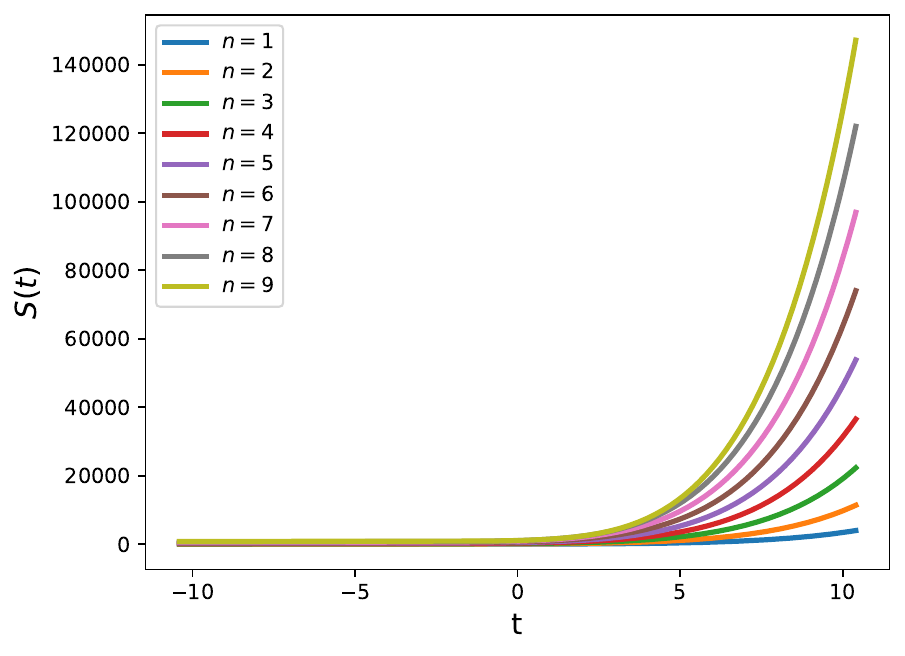}
		}
		\subfloat[\label{dS2bb}][]{%
			\includegraphics[width=0.4\textwidth]{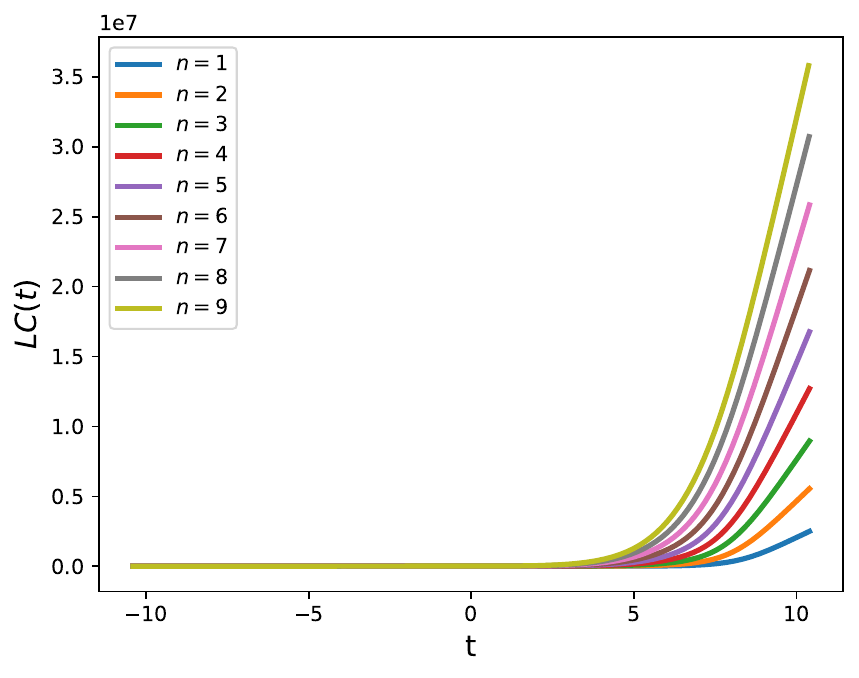}
		}
		
		\caption{Evolution of (a) entanglement entropy $S(t)$ and (b) Log classicality $LC(t)$ for de-Sitter expansion \eqref{eq:dsa} in $(3+1)-$dimensions. Here, $N=10$, $a_0=1$, $H=0.5$ and we count up to $l=200$.}
		\label{fig:dS2}
	\end{center}
\end{figure*}

For the de-Sitter expansion in $(3+1)-$dimensions, the normal modes \eqref{eq:nm3d} will the following form:
\begin{equation}
    \omega_{j,l}^2(t)=\frac{-9H^2}{4}+\frac{F_j^2(l)}{a^2(t)}\quad;\quad a(t)=a_0e^{H(t-t_0)}
\end{equation}
During a de-Sitter expansion \eqref{eq:dsa} in $(3+1)-$dimensions, we see from \ref{fig:dS2} that both the entanglement entropy and log classicality of all subsystem sizes exhibit unbounded growth in time, thereby fulfilling the classicality criteria at late-times. This is in stark contrast with the behaviour observed in $(1+1)-$dimensions, where the entanglement entropy remained constant, thereby failing the classicality criteria at late-times. The fluctuations therefore undergo a quantum-classical transition in a de-Sitter background in $(3+1)-$dimensions, but not in $(1+1)-$dimensions.

Physically, this implies that during a de-Sitter expansion in $(3+1)-$dimensions, the momentum modes of discretized linear fluctuations become inverted as they cross the Hubble radius. This inversion in turn causes \textit{spatial} bipartitions of fluctuations in the co-moving frame to both quickly decohere ($S\to\infty$) and also exhibit a high-degree of classical correlations ($LC\to\infty$). The Gaussian nature of fluctuations further enables Hermitian observables of the form in \eqref{eq:weyl} to be fully described in terms of two-point functions. However, non-trivial quantum signatures in such observables are rapidly suppressed in the classicality limit as discussed in $\eqref{eq:commratio}$. As a result, at late-times, real-space bipartitions of fluctuations in the co-moving frame are essentially described by classical statistical ensembles, with their phase-space distribution sharply peaking about classical trajectories. This also implies that at late-times, it is nearly impossible to distinguish whether these fluctuations were of quantum or classical origin without high-precision observations. However, the de-Sitter expansion is expected to have occurred only for a finite time ($\mathcal{N}\sim60$ e-folds) before it transitioned to a power-law expansion in the radiation-dominated epoch. In the next subsection, we will see how this transition impacts the classicality criteria.

\subsection{Transition from de-Sitter to radiation-dominated era}
In order to model the exit of inflation to radiation dominated era in the early Universe, we study the asymptotic results for two different types of transition --- i) a hard transition that gives us some analytic control, and ii) a smooth transition that can only be probed numerically.

\subsubsection{Hard Transition}
Let us consider the following scale-factor of expansion where the exit of inflation occurs at $t=t_e$:
\begin{equation}
    a(t)= \begin{cases}
        a_ee^{H_0(t-t_e)} & t\leq t_e \\
        a_e\sqrt{1+2H_0(t-t_e)} & t\geq t_e
    \end{cases}\quad;\quad a_e=a_0e^{H_0(t_0-t_e)},
\end{equation}
where $a_0$ and $a_e$ are the scale-factor values at initial time $t_0$ and transition time $t_e$ respectively, and $H_0$ is the Hubble constant during the de-Sitter expansion. For the above scale factor, the Hubble parameter ($H=\dot{a}/a$) is continuous at $t_e$, whereas the double derivative $\ddot{a}/a$ is not:
\begin{equation}
    \lim_{t\to t_e^{-}}H(t)=H_0=\lim_{t\to t_e^{+}}H(t)\quad;\quad \lim_{t\to t_e^{-}}\frac{\ddot{a}(t)}{a(t)}=-H_0^2\neq H_0^2 =\lim_{t\to t_e^{+}}\frac{\ddot{a}(t)}{a(t)}
\end{equation}
Therefore, the normal modes \eqref{eq:nm3d} are also discontinuous at $t=t_e$:
\begin{equation}\label{eq:whard}
    \omega_{j,l}^2(t)= \begin{cases}
        -\frac{9H_0^2}{4}+\frac{F_j^2(l)e^{-2H_0(t-t_e)}}{a_e^2} & t\leq t_e \\
        \frac{3H_0^2}{4\left(1+2H_0(t-t_e)\right)^2} + \frac{F_j^2(l)}{a_e^2\left(a+2H_0(t-t_e)\right)} & t>t_e
    \end{cases}.
\end{equation}
Upon imposing the continuity of the wave-function at $t=t_e$ through $b$ and $\dot{b}$ as proposed in \cite{2023Boutivas.etal}, we obtain the following late-time behavior for scaling parameters using \eqref{eq:nm3d} and \eqref{eq:bsolgen}:
\begin{equation}\label{eq:hardb}
    b_{j,l}^2(t\gg t_e)\sim \frac{a(t)}{a_0}\left(\frac{a_e^2H_0^2}{F_j^2(l)}\right)\left[\cos{\left[\zeta(t)t\right]}-\frac{2a_eH_0}{F_j(l)}\sin{\left[\zeta(t)t\right]}\right]^2\quad;\quad \zeta(t)=\frac{F_j(l)}{a_eH_0t}\left(\frac{a(t)}{a_e}-1\right)
\end{equation}
The above form for the scaling parameter indicates an oscillatory behaviour with decreasing frequency ($\zeta \sim t^{-1/2}$) and increasing amplitude ($b\propto t^{1/4}$) as the expansion proceeds. Similarly, the scale-factor at exit ($a_e$) increases exponentially with the number of e-folds of inflation ($a_e=a_0e^{\mathcal{N}}$), which in turn increases the amplitude of the oscillations ($b\propto a_e^2$) while damping the frequency ($\zeta\propto a_e^{-1}$). Since this expression holds at late-times, we expect these properties to be carried over even when considering a more realistic scenario of a smooth transition from de-Sitter to power-law (radiation-dominated) expansion. In the next subsection, we will therefore see how these features of scaling parameters $b(t)$ can dictate entanglement evolution even for a smooth transition.

\begin{figure*}[!hbt]
	\centering
	\includegraphics[scale=0.5]{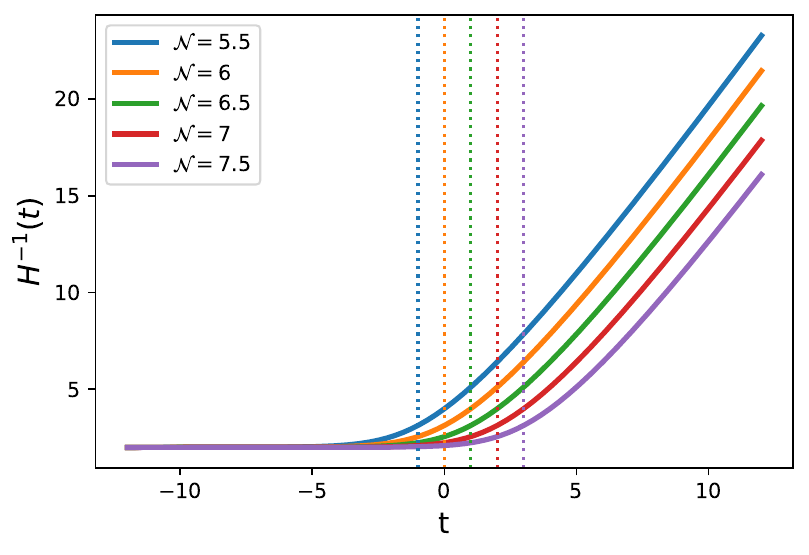}
	\caption{Evolution of Hubble radius $H^{-1}(t)$ for a smooth transition from de-Sitter expansion to radiation-dominated epoch~\eqref{eq:smoothtrans}, for various e-fold values ($\mathcal{N}=H_0(t_e-t_0)$). Here $H_0=0.5$, $t_0=-12$ and the vertical lines mark the corresponding values of $t_e$ \eqref{eq:smoothtrans}.}
		\label{fig:trans1}

\end{figure*}

\subsubsection{Smooth Transition}
To model a smooth transition from inflation to radiation dominated era of expansion, we look at the following functional form of Hubble parameter motivated in \cite{1998Gunzig.etalClassicalandQuantumGravity}:
\begin{equation}
    H=\frac{H_0}{1+\frac{a^2(t)}{a_e^2}}\quad;\quad a_e=a_0e^{H_0(t_e-t_0)},
\end{equation}
where $H_0$ is the Hubble parameter during inflation and $t_e$ denotes the end of inflation. The number of e-folds of inflation is given by $\mathcal{N}=\log{\frac{a_e}{a_0}}=H_0(t_e-t_0)$. On integrating the above equation, we get the corresponding scale-factor for the overall evolution:
\begin{equation}\label{eq:smoothtrans}
    a(t)=a_e\sqrt{\mathcal{W}_0\left[e^{2H_0(t-t_{e})}\right]} \sim\begin{cases}
        a_ee^{H_0(t-t_e)} & t\ll t_e \\
        a_e\sqrt{2H_0(t-t_e)} & t\gg t_e
    \end{cases},
\end{equation}
where $\mathcal{W}_0$ is the principal branch of Lambert W function. The above form of $a(t)$ makes it difficult to solve the Ermakov equation exactly. However, at late-times, we expect the scaling parameters to have a similar behaviour as obtained in \eqref{eq:hardb} for the hard transition.

\begin{figure*}[!ht]
	\begin{center}
		\subfloat[\label{trans2a}][]{%
			\includegraphics[width=0.4\textwidth]{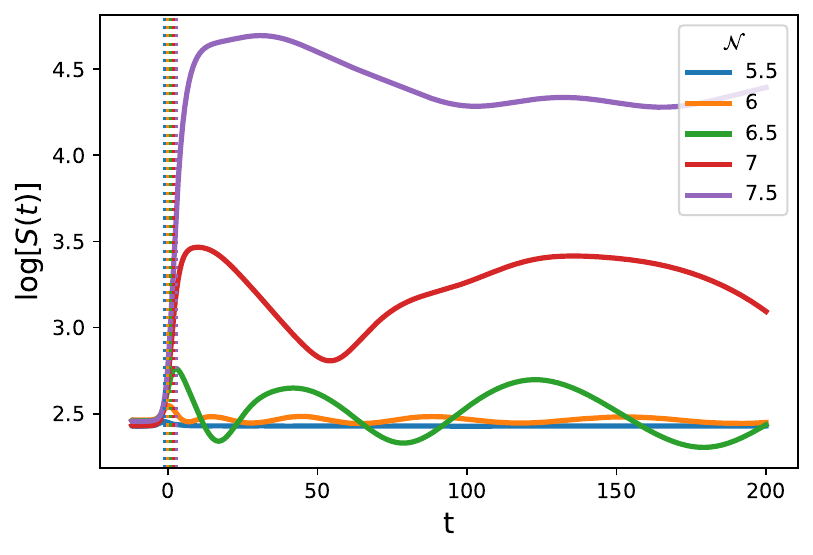}
		}
		\subfloat[\label{trans2b}][]{%
			\includegraphics[width=0.4\textwidth]{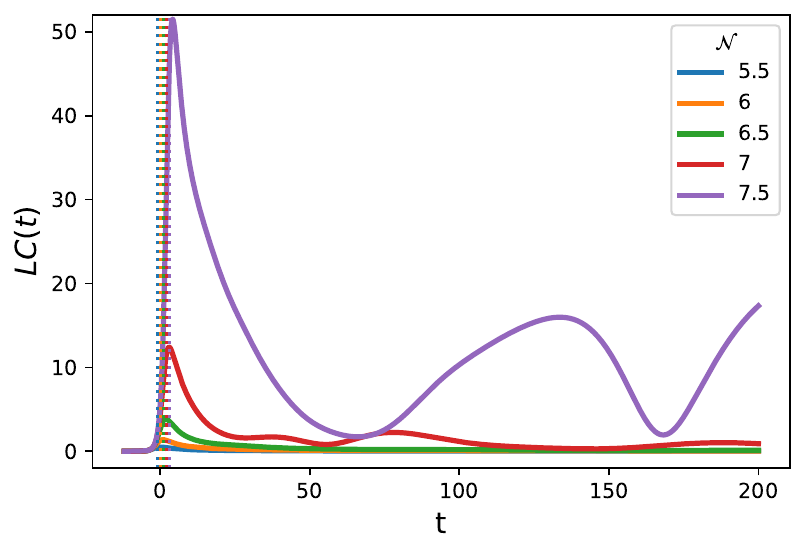}
		}
		
		\caption{Evolution of (a) entanglement entropy in log-scale, and (b) log-classicality for a smooth transition from inflation to radiation dominated epoch \eqref{eq:smoothtrans} for various e-fold values ($\mathcal{N}$) considered in \ref{fig:trans1}. Here, $n=1$, $N=5$, $t_0=-12$, $a_0=e^{-6}$, $H_0=0.5$ and we count up to $l=100$.}
		\label{fig:trans2}
	\end{center}
\end{figure*}

For this model, we test the classicality criteria for different e-fold values of inflation considered in \ref{fig:trans1}. From \ref{fig:trans2}, we observe that the runaway growth in entanglement entropy and log classicality is cut off as inflation ends, transitioning to an oscillatory behaviour in the radiation-dominated era. We also observe that the entanglement entropy and log classicality growths are highly sensitive to the number of e-folds $\mathcal{N}$, whereas the oscillation frequency decays with $\mathcal{N}$, matching the features of scaling parameters in \eqref{eq:hardb}. It can also be seen that, just as in \eqref{eq:hardb}, the oscillation frequency decreases as the expansion proceeds in the radiation-dominated era. We may therefore infer that there is a runaway classicalization during the inflationary phase (marked by an irreversible loss in quantum coherence) after which the spatial subregions proceed to retain remnant quantum signatures.

While it may be computationally demanding to simulate $\mathcal{N}\sim 60$ e-folds of inflation using larger system sizes ($N\gg1$), we expect the classicalization of spatial subregions to be further exacerbated upon scaling up these parameters. The leading-order classical behaviour arising from longer e-folds of inflation is therefore expected to continue into the radiation-dominated era as well, and can in principle facilitate an equivalent description via stochastic fluctuations.

\section{Conclusions and Discussions}\label{sec:conc}

In this work, our primary focus is to understand the quantum-to-classical transition of entangled quadratic systems with spatial degrees of freedom. Our investigation involved three distinct signatures of classical behavior: i) decoherence as a measure of how well the system can be described by a classical statistical ensemble, ii) runaway squeezing of the Wigner function about classical phase-space trajectories, and iii) rapid suppression of non-commutativity in observables. We developed the necessary tools in Section \ref{sec:cho} to extract and measure these signatures in terms of entanglement entropy $S(t)$, log classicality $LC(t)$, and relative strength $R_{(x,p)}$ from a multi-mode Gaussian state.

We obtained a simple geometric picture of the interplay between these signatures through the stability analysis of the reduced Wigner function of the subsystem, as illustrated in \ref{fig:Wigfun}. The results, summarized in Table \ref{tab:CHO}, reveals that the presence of instabilities arising from a gapped inverted mode spectrum in the system leads to the emergence of all three classicality signatures in the CHO (quadratic system). On the other hand, other stability regimes exhibited only partial or no indications of classical behavior.

In Section \ref{sec:1dcosmology}, we analyzed linear fluctuations of an expanding background in $(1+1)$ dimensions. We found that a quantum-to-classical transition did not occur as the dynamics preempted decoherence. This was demonstrated by considering two different scale factors of expansion: i) a Tanh expansion with fixed values at asymptotic past and future, and ii) an exponentially growing scale factor corresponding to a de-Sitter expansion. 
In Section \ref{sec:3dcosmology}, we extended the analysis to $(3+1)-$dimensions 
and showed that for Tanh expansion the fluctuations failed to classicalize, whereas for a de-Sitter expansion the fluctuations underwent a quantum-to-classical transition. We further showed that this transition is cut off when inflation ends and the background proceeds to power-law (radiation dominated) expansion. However, the leading order classical behaviour of the fluctuations arising from the inflationary epoch appeared to be irreversible at late-times.


Throughout, we discovered that the inversion of normal modes in momentum-space acted as a common trigger for the emergence of classical behavior. While this inversion had limited impact on the momentum space of quadratic systems, it significantly affected the entangled degrees of freedom in real-space.  For a flat background in $(1+1)$ dimensions, recent studies have revealed that: i) the entanglement entropy ``classicalizes" i.e., it mimics the statistical entropy of classically chaotic systems via a linear growth, wherein the growth rate is given by the sum of all positive Lyapunov exponents~\cite{2018Bianchi.etalJournalofHighEnergyPhysics,2018Hackl.etalPhys.Rev.A,2023Chandran.ShankaranarayananPhys.Rev.D,2023Boutivas.etalJournalofHighEnergyPhysics}, 
ii) the leading order behavior of entanglement entropy asymptotically converges with other correlation measures, such as fidelity, Loschmidt echo, and circuit complexity of the entire system~\cite{2023Chandran.ShankaranarayananPhys.Rev.D}, and iii)  entanglement entropy asymptotically transitions from an area-law to a volume-law with subsystem size, thereby mimicking thermodynamic entropy~\cite{2023Katsinis.etal,2022Bianchi.etalPRXQuantum,2023Chandran.ShankaranarayananPhys.Rev.D,2023Boutivas.etal}. In addition to our analysis, these effects further signal the emergence of both classical \textit{and} possible thermodynamic behaviour in the real-space from quantum foundations. However, the exact generalization of these properties to higher dimensions is subject of future work and will be addressed elsewhere using the tools developed here.

The computational limitations in managing exponentially growing scaling parameters due to mode inversion are much more pronounced when simulating large system-sizes. While boundary effects are a cause for concern when studying small-system sizes, earlier works have shown that the the IR cutoff ($N\tilde{d}$) dependence of entanglement entropy is typically suppressed in the energy scales of interest~\cite{2005Casini.HuertaJ.Stat.Mech.,2021Jain.etalPhys.Rev.D}. We therefore do not expect it to play a major role in the quantum-to-classical transition problem. However, the IR-terms in entanglement entropy and log classicality need to be rigorously investigated for more insight into the matter, which we hope to address in the future.

Since the temperature fluctuations in the CMB are predominantly Gaussian as per current observations, the Gaussian state we have considered here is sufficient for addressing the quantum-to-classical transition of early-universe fluctuations. However, it is not currently understood whether the presence of non-Gaussianities would accelerate or slow down classicalization, and therefore requires further investigation. Resolving this can filter out the spatial effects exclusively arising from non-Gaussianity in the context of the quantum-to-classical transition problem while also laying out potential new ways of obtaining direct observational evidence for the quantum origin of CMB fluctuations.

Our analysis further provides the tools necessary to distinguish between cosmological models, such as those with similar observable power spectra, as has been the subject of recent investigations~\cite{2023Raveendran.Chakraborty}. Of particular interest is using these measures to distinguish inflation from bounce, which is currently under investigation. Lastly, a generalization of our real-space approach to account for higher-order curvature perturbations is an outstanding problem we hope to address in future works.


\begin{acknowledgements}
The authors thank Suddhasattwa Brahma, Ashu Kushwaha, Orlando Luongo, Amaury Micheli, Abhishek Naskar and Vincent Vennin for useful discussions. The authors also thank the referee for their invaluable comments. SMC is supported by Prime Minister's Research Fellowship offered by the Ministry of Education, Govt. of India. The work is supported by the SERB Core Research grant.
\end{acknowledgements}
\appendix

\section{Conditions for purity saturation}\label{app:saturation}
The purity of CHO has the following form:
\begin{equation}
        \delta_{QD}(t)=\sqrt{\frac{4K_+K_-}{(K_++K_-)^2+(L_+-L_-)^2}}\quad;\quad K_\pm=\frac{\omega_\pm(t_0)}{b_\pm^2}\quad;\quad L_\pm=\frac{\dot{b}_\pm}{b_\pm}
    \end{equation}

Let us now rewrite $b_-(t)=f(t)b_+(t)$:
\begin{equation}
    \delta_{QD}(t)=\sqrt{\frac{4\omega_+(t_0)\omega_-(t_0)}{(f\omega_+(t_0)+f^{-1}\omega_-(t_0))^2+\dot{f}b_+^4}}\xRightarrow[]{\dot{f}\to0}\frac{\sqrt{4\omega_+(t_0)\omega_-(t_0)}}{f\omega_+(t_0)+f^{-1}\omega_-(t_0)},
\end{equation}
where we see that the evolution of purity, and in turn, entanglement entropy \eqref{eq:CHO-ent1} saturates in regimes where the Ermakov solutions $b_\pm(t)$ have the same time-evolution ($\dot{f}=0$) upto a proportionality constant ($f$).

\section{Entanglement entropy of CHO}\label{app:cho}
Like in the case of time-independent CHO~\cite{1993-Srednicki-Phys.Rev.Lett.,2006Ahmadi.etalCan.J.Phys.}, to evaluate the entanglement entropy, we must first calculate the eigenvalues of the reduced density matrix (RDM) of the system~\cite{1993-Srednicki-Phys.Rev.Lett.,2017Ghosh.etalEPLEurophysicsLetters} by solving the following integral equation~\cite{1993-Srednicki-Phys.Rev.Lett.,2006Ahmadi.etalCan.J.Phys.}:
\begin{equation}
	\int dx'_2 \, \rho_2(x_2,x_2')f_n(x_2')=p_nf_n(x_2) \, .
\end{equation}
The solution for the above integral equation is~\cite{2017Ghosh.etalEPLEurophysicsLetters}:
\begin{align}
	f_n(x)&=\frac{1}{\sqrt{2^n n!}}\left(\frac{\epsilon}{\pi}\right)^{1/4}H_n(\sqrt{\epsilon}x)\exp{-\left(\epsilon+i\delta\right)\frac{x^2}{2}}\nonumber\\
	\epsilon&=\sqrt{\Gamma_1^2-\Gamma_2^2}\nonumber\\
	p_n&=\left(1-\xi(t)\right)\xi^n(t)\\
	\xi(t)&=\frac{\Gamma_2}{\Gamma_1+\epsilon}=\frac{\sqrt{\left(\frac{\omega_+(t_0)}{b_+^2(t)}+\frac{\omega_-(t_0)}{b_-^2(t)}\right)^2+\left(\frac{\dot{b}_+(t)}{b_+(t)}-\frac{\dot{b}_-(t)}{b_-(t)}\right)^2}-2\sqrt{\frac{\omega_+(t_0)\omega_-(t_0)}{b_+(t)b_-(t)}}}{\sqrt{\left(\frac{\omega_+(t_0)}{b_+^2(t)}+\frac{\omega_-(t_0)}{b_-^2(t)}\right)^2+\left(\frac{\dot{b}_+(t)}{b_+(t)}-\frac{\dot{b}_-(t)}{b_-(t)}\right)^2}+2\sqrt{\frac{\omega_+(t_0)\omega_-(t_0)}{b_+(t)b_-(t)}}} \nonumber
\end{align}

The entanglement entropy is calculated as follows: 
\begin{equation}
\label{eq:CHO-ent1}
	S(t)=-\sum_n p_n\log{p_n}=-\log{[1-\xi(t)]}-\frac{\xi(t)}{1-\xi(t)}\log{\xi(t)},
\end{equation}
\section{Phase space stability analysis of CHO}\label{sec:ps}
The vacuum states are typically well-defined when the Hamiltonian becomes time-independent. Hence in the case of CHO we consider an evolution in $\omega(t)$ and $\chi(t)$ that are asymptotically constant. In Ref. \cite{2023Chandran.ShankaranarayananPhys.Rev.D}, the authors showed that the asymptotic values of the normal modes decided the late-time stability of the system, the signatures of which were obtained from various correlation measures. Similarly, we may consider the stability analysis of the quantum state in the phase-space via the Wigner function. Let us set the values of the two normal modes --- $\omega_+^2(t)=\omega^2(t)$ and $\omega_-^2(t)=\omega^2(t)+2\chi^2(t)$ --- to constant values $u_+^2$ and $u_-^2$ ($u_+^2 \leq u_-^2$), respectively at late-times. In the asymptotic future ($t\to\infty$), the Ermakov equation, therefore, takes the following form: 
\begin{equation}
	\ddot{b}_j(t)+u_j^2b_j(t) \sim \frac{\omega_j^2(t_0)}{b_j^3(t)}\quad;\quad j=+,-
\end{equation}
Since the co-efficient in the second term of the above equation is time-independent, we can obtain the following solutions~\cite{2017Ghosh.etalEPLEurophysicsLetters}:
\begin{equation}
\label{eq:bjlongtime}
	b_j(t)\sim \sqrt{1+\left(\frac{\omega_j^2(t_0)}{u_j^2}-1\right)\sin^2{u_jt}}\quad;\quad \dot{b}_j(t)\sim \left(\omega_j^2(t_0)-u_j^2\right)\frac{\sin{2u_jt}}{2u_jb_j(t)} \, .
\end{equation}
We now look at various stability regimes of these solutions below and track its features in the phase-space picture (see \ref{fig:Wigfun}) :
\begin{itemize}
    \item Stable Modes $u_{j}^2>0$ : Scaling parameters $\{b_{j}\}$ are oscillatory and bounded.
    \item Zero Modes $u_j^2=0$ :
            \begin{equation}\label{b:zero}
		b_j(t)\sim \omega_j(t_0)t\quad;\quad \dot{b}_j(t)\sim \omega_j(t_0)
	\end{equation}
 Suppose $\omega_+$ is a zero mode and $\omega_-$ is a stable mode. At late-times, we have:
 \begin{equation}
     \delta_{QD}\sim \frac{2}{t}\sqrt{\frac{K_-}{(K_-^2+L_-^2)\omega_+(t_0)}}\quad;\quad \mathscr{C}\sim \frac{1}{\sqrt{1+\frac{K_-^2+L_-^2}{K_-\omega_+(t_0)}}}
 \end{equation}
 We see that the purity falls to zero as $t\to\infty$, whereas classicality parameter retains its oscillatory behaviour about a value between 0 and 1, i.e., there is no runaway squeezing.


    \item Inverted Modes $u_j^2<0$ : At late times, the solutions \eqref{eq:bjlongtime} further reduce to :
	\begin{equation}\label{b:inverted}
		b_j(t)\sim c_je^{v_jt}\quad;\quad \dot{b}_j(t)\sim c_jv_je^{v_jt} \quad;\quad c_j=\frac{1}{2}\sqrt{1+\frac{\omega_j^2(t_0)}{v_j^2}}
	\end{equation}
where we have defined $u_j=iv_j$. When both modes are inverted, we see that $v_+\geq v_-$ in general, and as a result:
\begin{equation}
    \lim_{t\to\infty}\delta_{QD}\sim \begin{cases} \frac{2\sqrt{\omega_+(t_0)\omega_-(t_0)}}{c_+c_-(v_+-v_-)}e^{-(v_++v_-)t} &v_+>v_-\\\frac{2c_+c_-\sqrt{\omega_+(t_0)\omega_-(t_0)}}{c_+^2\omega_-(t_0)+c_-^2\omega_+(t_0)} & v_+\to v_-    
    \end{cases}
\end{equation}
The result gives us two distinct cases --- if $v_+>v_-$ (gapped), the long-time limit will always result in a purity that exponentially decays to zero, thereby exhibiting rapid decoherence. On the other hand, if the inverted modes converge asymptotically (i.e., ungapped), the subsystem is protected from further decoherence (this conditions for purity saturation is much more general, as worked out in Appendix \ref{app:saturation}). The degree of classical correlation $\delta_{CC}$, on the other hand, has the following late-time behaviour:
\begin{equation}
    \lim_{t\to\infty}\mathscr{C}^2\sim \begin{cases} 1-\frac{\omega_+(t_0)c_-^2}{\omega_-(t_0)c_+^2}\left(1-\frac{v_-}{v_+}\right)^2 e^{-2(v_+-v_-)t} &v_+>v_-\\1-\frac{\omega_+(t_0)\omega_-(t_0)}{c_+^2c_-^2v^2}e^{-4vt} & v_+\to v_-\sim v  
    \end{cases}
\end{equation}
We see that at late times $\mathscr{C}\to1$, with the squeezing being much faster in the ungapped case than in the gapped case. Therefore, we see that the only case that simultaneously results in both rapid decoherence and runaway squeezing, thereby satisfying the classicality criteria, is when the system develops gapped inverted modes.
\end{itemize}

\section{Particle production at late-times due to instabilities}\label{app:pp}
In order to quantify particle production due to such instabilities, it is essential to specify the `in'-states and `out'-states with respect to which ladder operators for each $k-$mode are defined:
\begin{equation}
    \mathscr{N}_k=\bra{in}a_k^{\dagger(out)}a_k^{(out)}\ket{in}=\abs{\beta}^2\quad;\quad a_k^{(out)}=\alpha a_k^{(in)}+\beta a_k^{\dagger(in)},
\end{equation}
where the ladder operators are related via a Bogoliubov transformation. Invoking the mode-evolution in \eqref{GS}, we may consider the vacuum state at time $t_0$ as the `in'-state, and the evolution to a later time $t$ as the `out'-state. The ladder operators for the `out'-state can then be described as follows~\cite{2008LoheJournalofPhysicsAMathematicalandTheoretical}:
\begin{equation}
    a_k^{(out)}\ket{out}=0\quad;\quad a_k^{(out)}=\frac{e^{i\omega(t_0)\tau}}{\sqrt{2}}\left\{\left(\frac{\sqrt{\omega_k(t_0)}}{b_k}-\frac{i\dot{b_k}}{\sqrt{\omega_k(t_0)}}\right)x+\frac{ib_k}{\sqrt{\omega_k(t_0)}}p\right\}.
\end{equation}
The coefficients satisfying the transformation are obtained below:
\begin{equation}
    \alpha=\frac{e^{i\omega_k(t_0)\tau}}{2}\left\{\frac{1}{b_k}+b_k-\frac{i\dot{b}_k}{\omega_k(t_0)}\right\}\quad;\quad \beta=\frac{e^{i\omega_k(t_0)\tau}}{2}\left\{\frac{1}{b_k}-b_k-\frac{i\dot{b}_k}{\omega_k(t_0)}\right\}
\end{equation}
The particle number expectation for the $k-$mode can then be obtained as follows:
\begin{equation}
    \mathscr{N}_k=\frac{1}{4}\left\{\left(\frac{1}{b_k}-b_k\right)^2+\frac{\dot{b}_k^2}{\omega_k^2(t_0)}\right\}
\end{equation}

Let us now perform a stability analysis of the $k-$mode at late-times as done in Appendix \ref{sec:ps}, and see how it affects particle production:
\begin{itemize}
    \item If the mode evolves to a zero-mode at late-times, we have $b_k\sim \omega_k(t_0)t$.:
    \begin{equation}
        \mathscr{N}_k\sim \frac{\omega_k^2(t_0)t^2}{4}
    \end{equation}
    \item If the mode is inverted at late-times ($\lim_{t\to\infty}\omega_k\to iv_k$), we have $b_k \propto e^{v_k t}$:
    \begin{equation}
        \mathscr{N}_k\sim \left[\frac{v_k}{\omega_k(t_0)}+\frac{\omega_k(t_0)}{v_k}\right]^2\frac{e^{2v_kt}}{16}
    \end{equation}
\end{itemize}
In both the above cases, we see that particle production with respect to the vacuum (`in'-state) becomes unbounded due to instabilities persisting at late-times. 

\subsection*{Covariance matrix and particle number}
The covariance matrix for each $k-$mode will take the following form:
\begin{equation}
    \langle \{\hat{x}_k,\hat{x}_k\}\rangle=\frac{b_k^2}{\omega_k(t_0)}\quad;\quad \langle \{\hat{p}_k,\hat{p}_k\}\rangle=\frac{\omega_k(t_0)}{b_k^2}+\frac{\dot{b}_k^2}{\omega_k(t_0)}\quad;\quad \langle \{\hat{x}_k,\hat{p}_k\}\rangle=\frac{b_k\dot{b}_k}{\omega_k(t_0)}
\end{equation}
Since the $k-$mode is in a pure state, the entanglement entropy calculated from the covariance matrix is trivially zero. However, the log classicality takes the following form:
\begin{equation}
    LC=\log{\sqrt{1+\frac{b_k^2\dot{b}_k^2}{\omega_k^2(t_0)}}}
\end{equation}
The stability analysis of log classicality leads to the following relation at late-times:
\begin{equation}
    LC\sim \begin{cases}
        \frac{1}{2}\log{\mathscr{N}_k} & \quad\text{Zero mode}\\
        \log{\mathscr{N}_k} & \quad \text{Inverted mode}
    \end{cases}
\end{equation}

We see that for each $k-$mode, the extent of particle production, despite being a purely quantum phenomenon, is reflected in the log classicality measure that captures the build-up of classical phase-space correlations from Wigner function. While obtaining a similar expression for spatial subsystems is beyond the scope of this work, we expect the final relation to also include entanglement entropy. A similar result involving particle number, squeezing parameter, and purity for a subsystem interacting with environmental degrees of freedom was obtained in \cite{2022Martin.etalJCAP}.

\section{Classicality criteria and Canonical transformations}\label{sec:dss}
In Section \ref{sec:cho} we were able to successfully extend Morikawa's classicality criteria to multi-mode Gaussian states. In this section, we show that it is however not completely independent of the choice of conjugate variables. 

\subsection*{Time-dependent Harmonic Oscillator}
To investigate the effects of canonical transformations, let us consider the Hamiltonian of a time-dependent oscillator as follows:
\begin{equation}
    \mathscr{H}^{(I)}(\eta)=\frac{P^2}{2}+\frac{\omega_{_{I}}^2(\eta)X^2}{2}=\frac{P^2}{2}+\frac{a^2(\eta)\Omega^2(\eta)X^2}{2} \,  ,
\end{equation}
where we now use $\eta$ as the time coordinate for comparison. The wave-function that describes the system is a solution to the time-dependent Schr\"{o}dinger equation, and unitarily evolves from an initial state defined at $\eta=\eta_0$ as follows~\cite{1967LewisPhys.Rev.Lett.,2009CampbellJ.Phys.A.}:
\begin{equation}
    \Psi(\eta)=\exp{-i\int_{\eta_0}^\eta \mathscr{H}^{(I)}(\eta')d\eta'}\Psi(\eta_0)
\end{equation}
Let us now transform the Hamiltonian $\mathscr{H}^{(I)}(\eta) \to \mathscr{H}^{(II)}(\eta)$ as follows:
\begin{equation}\label{eq:hamconnection}
    \mathscr{H}^{(II)}(t)=\frac{\mathscr{H}^{(I)}(\eta(t))}{a(\eta(t))}=\frac{P^2}{2a(\eta(t))}+\frac{a(\eta(t))\Omega^2(\eta(t))X^2}{2}
\end{equation}
With the above rescaling, the time-evolution of a particular state can be preserved by also rescaling the time-coordinate appropriately:
\begin{equation}
    \int_{\eta_0}^\eta \mathscr{H}^{(I)}(\eta)d\eta=\int_{t_0}^t \mathscr{H}^{(II)}(t) dt\quad;\quad t=\int a(\eta)d\eta
\end{equation}
Now, we employ the following canonical transformations with respect to $\mathscr{H}^{(II)}$~\cite{1987PedrosaJournalofMathematicalPhysics}:
\begin{equation}\label{eq:cantrans}
    X=\frac{x}{\sqrt{a(t)}} \quad;\quad P=\sqrt{a(t)}p-\frac{\dot{a}(t)}{2\sqrt{a(t)}}x
\end{equation}
The resultant Hamiltonian is:
\begin{equation}\label{eq:frequencyrelation}
    \mathscr{H}^{(II)}(t)=\frac{p^2}{2}+\frac{\omega_{II}^2(t)x^2}{2}\quad;\quad \omega_{II}^2(t)=\frac{\omega_{I}^2(\eta(t))}{a^2(t)}+\frac{1}{4}\left(\frac{\dot{a}(t)}{a(t)}\right)^2-\frac{\ddot{a}(t)}{2a(t)}
\end{equation}
We now look at how the scaling parameters corresponding to $\mathscr{H}^{(I)}$ and $\mathscr{H}^{(II)}$, namely $B(\eta)$ and $b(t)$ are related. For this, we look at the non-linear Ermakov equation:

\begin{equation}\label{eq:ermakovv}
	B''(\eta)+\omega_{I}^2(\eta)B(\eta)=\frac{\omega_{I}^2(\eta_0)}{B^3(\eta)}
\end{equation}
To arrive at a solution for the Ermakov equation, we first consider solutions to the classical time-dependent oscillator:
\begin{equation}
    Y''(\eta)+\omega_{I}^2(\eta)Y(\eta)=0
\end{equation}
From a set of independent solutions $Y_1(\eta)$ and $Y_2(\eta)$ of the above equation, the scaling parameter $B(\eta)$ can be obtained as follows:
\begin{multline}\label{eq:bsolgen}
    B^2(\eta)=\frac{B^2(\eta_0)}{W_{Y}^2}\left\{Y_1(\eta)Y'_2(\eta_0)-Y_1'(\eta_0)Y_2(\eta)+\frac{B'(\eta_0)}{B(\eta_0)}\left(Y_1(\eta)Y_2(\eta_0)-Y_2(\eta)Y_1(\eta_0)\right)\right\}^2\\+\frac{\omega_{I}^2(\eta_0)}{W_{Y}^2B^2(\eta_0)}\left\{Y_1(\eta)Y_2(\eta_0)-Y_2(\eta)Y_1(\eta_0)\right\}^2,
\end{multline}
where $W_Y$ is the Wronskian for solutions $Y_1(\eta)$ and $Y_2(\eta)$. On imposing the initial conditions $B(\eta_0)=1$ and $B'(\eta_0)=0$, we get:
\begin{equation}\label{eq:bsolution}
    B^2(\eta)=\frac{1}{W_{Y}^2}\left[\left\{Y_1(\eta)Y'_2(\eta_0)-Y_1'(\eta_0)Y_2(\eta)\right\}^2+\omega_{I}^2(\eta_0)\left\{Y_1(\eta)Y_2(\eta_0)-Y_2(\eta)Y_1(\eta_0)\right\}^2\right].
\end{equation}
Similarly, for Hamiltonian $\mathscr{H}^{(II)}$, we write down the classical equation of motion and Ermakov equations respectively as follows:
\begin{equation}
    \ddot{y}(t)+\omega_{II}^2(t)y(t)=0\quad;\quad \ddot{b}(t)+\omega_{II}^2(t)b(t)=\frac{\omega_{II}^2(t_0)}{b^3(t)}
\end{equation}
Suppose the independent solutions are $y_1(t)$ and $y_2(t)$, the scaling parameter $b(t)$ are obtained as follows:
\begin{equation}\label{eq:bgeneral}
    b^2(t)=\frac{1}{W_y^2}\left[\left\{y_1(t)\dot{y}_2(t_0)-\dot{y}_1(t_0)y_2(t)\right\}^2+\omega_{II}^2(t_0)\left\{y_1(t)y_2(t_0)-y_2(t)y_1(t_0)\right\}^2\right],
\end{equation}
where $W_y$ is the Wronskian for solutions $y_1(t)$ and $y_2(t)$. The above solution automatically satisfies the initial conditions $b(t_0)=1$ and $\dot{b}(t_0)=0$. Using the equation connecting frequencies $\omega_I^2(t)$ and $\omega_{II}^2(\eta)$ in \eqref{eq:frequencyrelation}, we obtain the following relations connecting $y(t)$ and $Y(\eta)$:
\begin{equation}
    y(t)=\sqrt{a(t)}Y(\eta(t))\quad;\quad \dot{y}(t)=\frac{1}{\sqrt{a(t)}}\left(Y'(\eta(t))+\frac{\dot{a}(t)}{2}Y(\eta(t))\right);\quad W_y=W_Y
\end{equation}
Substituting this back into the solution $b(t)$, we obtain the following relation:
\begin{multline}\label{eq:bconnection2}
    b^2(t)=\frac{a(t)}{a(t_0)}B^2(\eta)+\frac{a(t)a(t_0)}{2W_Y^2}\left\{\frac{\dot{a}^2(t_0)}{a^2(t_0)}-\frac{\ddot{a}(t_0)}{a(t_0)}\right\}\left[Y_1(\eta)Y_2(\eta_0)-Y_1(\eta_0)Y_2(\eta)\right]^2\\+\frac{a(t)\dot{a}(t_0)}{W_Y^2a(t_0)}\left[Y_1(\eta)Y_2'(\eta_0)-Y_1'(\eta_0)Y_2(\eta)\right]\left[Y_1(\eta)Y_2(\eta_0)-Y_1(\eta_0)Y_2(\eta)\right]
\end{multline}
The above expression relates the time-evolution from the respective vacuum states corresponding to $\mathscr{H}^{(I)}$ and $\mathscr{H}^{(II)}$. Alternatively, one may be interested in studying the evolution of, say, the $\eta-$vacuum in the $t$ representation (see, for instance, \cite{2018Rajeev.etalGeneralRelativityandGravitation}). This leads to a simplified relation between the corresponding scaling parameters. To see this, notice that the wave functions, in the two different representations, of a given state of the system are related via:
\begin{equation}
    \Psi_{II}(x,t)=\frac{1}{a^{1/4}(t)}\Psi_{I}[X(x),\eta(t)]\exp{i\frac{\dot{a}(t)}{4a(t)}x^2}
\end{equation}
For the special case of a Gaussian state, the above relation translates to the following relation between the corresponding scaling parameters:
\begin{equation}\label{eq:bconnection}
    \frac{\omega_{II}(t_0)}{b^2(t)}=\frac{\omega_{I}(\eta_0)}{a(t)B^2(\eta)}\quad;\quad \frac{\dot{b}(t)}{b(t)}=\frac{1}{a(t)}\left[\frac{B'(\eta(t))}{B(\eta(t))}+\frac{\dot{a}(t)}{2}\right]
\end{equation}
Consequently, the above relation is also valid if one can further specialize to the vacuum state of one of the representations. The relevance of this relation is that its direct extension to the case of harmonic lattices can be used to study the consequences of canonical transformations. Note that in the limit 
\begin{equation}\label{eq:adotinit}
    \dot{a}(t_0)\rightarrow0\quad\text{and}\quad \ddot{a}(t_0)a(t_0)\rightarrow 0
\end{equation} we obtain \eqref{eq:bconnection} from \eqref{eq:bconnection2}. This limit, therefore, translates to the case when the instantaneous vacua of both representations coincide at $t=t_0$.

\subsection*{Time-dependent CHO}

In order to observe the effects of canonical transformations on the classicality criteria, we now look at the CHO:
\begin{equation}
\mathscr{H}^{(I)}(\eta) =\frac{P_1^2}{2 }+\frac{P_2^2}{2 }+\frac{1}{2} \omega_I^2(\eta)\left(X_1^2+X_2^2\right)+\frac{1}{2}\chi_I^2(\eta)\left(X_1-X_2\right)^2\quad;\quad \mathscr{H}^{(II)}=\frac{\mathscr{H}^{(I)}[\eta(t)]}{a[\eta(t)]}
\end{equation}
The canonical transformations in \eqref{eq:cantrans} result in the following Hamiltonian:
\begin{align}
    \mathscr{H}^{(II)}(t)&=\frac{p_1^2}{2}+\frac{p_2^2}{2}+\frac{1}{2}\omega_{II}^2(t)(x_1^2+x_2^2)+\frac{1}{2}\chi_{II}^2(t)(x_1^2-x_2^2)\\ \omega_{II}^2(t)&=\frac{\omega_{I}^2[\eta(t)]}{a^2(t)}+\frac{1}{4}\left(\frac{\dot{a}(t)}{a(t)}\right)^2-\frac{\ddot{a}(t)}{2a(t)}\\ \chi_{II}^2(t)&=\frac{\chi_I^2[\eta(t)]}{a^2(t)}
\end{align}
In terms of $K_\pm$ and $L_\pm$ defined with respect to Hamiltonians $\mathscr{H}^{(I)}$ and $\mathscr{H}^{(II)}$ as given in \eqref{eq:rdmredef}, we get:
\begin{equation}
    K_\pm^{(II)}(t)=\frac{K_\pm^{(I)}[\eta(t)]}{a(t)}\quad;\quad L_\pm^{(II)}(t)=\frac{1}{a(t)}\left(L_\pm^{(I)}[\eta(t)]+\frac{\dot{a}(t)}{2}\right)
\end{equation}
We now look at how the characteristic parameters of the Wigner function are affected upon going from $\mathscr{H}^{(I)}$ described in terms of time $\eta$ to $\mathscr{H}^{(II)}$ described in terms of time $t$:
\begin{itemize}
    \item Degree of Quantum Decoherence $\delta_{QD}$ : 

    \begin{align}
        \delta_{QD}^{(II)}(t)&=\sqrt{\frac{4K_+^{(II)}(t)K_-^{(II)}(t)}{(K_+^{(II)}(t)+K_-^{(II)}(t))^2+(L_+^{(II)}(t)-L_-^{(II)}(t))^2}}\nonumber\\
        &=\sqrt{\frac{4K_+^{(I)}[\eta(t)]K_-^{(I)}[\eta(t)]}{(K_+^{(I)}[\eta(t)]+K_-^{(I)}[\eta(t)])^2+(L_+^{(I)}[\eta(t)]-L_-^{(I)}[\eta(t)])^2}}=\delta_{QD}^{(I)}[\eta(t)].
    \end{align}
\item Degree of Classical Correlation $\delta_{CC}$ : 
\begin{align}
        \frac{1}{\delta_{CC}^{(II)}(t)}&=\frac{K_+^{(II)}(t)L_-^{(II)}(t)+K_-^{(II)}(t)L_+^{(II)}(t)}{\sqrt{K_+^{(II)}(t)K_-^{(II)}(t)\left[(K_+^{(II)}(t)+K_-^{(II)}(t))^2+(L_+^{(II)}(t)-L_-^{(II)}(t))^2\right]}}\nonumber\\
        &=\frac{1}{\delta_{CC}^{(I)}(\eta)}\left[1+\left(\frac{\dot{a}(t)}{2}\right)\frac{K_+^{(I)}[\eta(t)]+K_-^{(I)}[\eta(t)]}{K_+^{(I)}[\eta(t)]L_-^{(I)}[\eta(t)]+K_-^{(I)}[\eta(t)]L_+^{(I)}[\eta(t)]}\right]
    \end{align}

\end{itemize}
Upon plugging the above expressions into \eqref{eq:dccc} and \eqref{eq:EntNHO}, we see that entanglement entropy (being a symplectic invariant~\cite{2003Serafini.etalJournalofPhysicsBAtomicMolecularandOpticalPhysics}) stays invariant under the canonical transformation in \eqref{eq:cantrans}, whereas log classicality does not:
\begin{equation}
    S^{(II)}(t)=S^{(I)}(\eta(t))\quad;\quad LC^{(II)}(t)\neq LC^{(I)}(\eta(t))\quad;\quad t=\int a(\eta)d\eta
\end{equation}
For the special case where $a(t)=a_0$ (constant), however, we see that they are both invariant~\cite{2023Chandran.ShankaranarayananPhys.Rev.D}:
\begin{equation}
    S^{(II)}(t)=S^{(I)}(a_0^{-1}t)\quad;\quad LC^{(II)}(t)= LC^{(I)}(a_0^{-1}t)\quad;\quad t= a_0 \eta
\end{equation}
The classicality criteria therefore has an ambiguity --- the condition on classicality parameter $\mathscr{C}$ is subject to change under a canonical transformation, even for the same time-evolved state, and the same subsystem division. Therefore, in order to manage this ambiguity, we make the second condition stronger by claiming that both representations $\mathscr{H}^{(I)}$ and $\mathscr{H}^{(II)}$ must satisfy the classicality criteria in \eqref{eq:criteria}, 
\begin{equation}
    \lim_{t\to\infty}S\to\infty \quad;\quad \lim_{t\to\infty} LC\to\infty,\nonumber
\end{equation}
failing which an asymptotic quantum-classical transition may be ruled out. 


%

\end{document}